\documentclass[review]{elsarticle}

\usepackage{lineno,hyperref}
\usepackage{comment,color}

\usepackage{hyperref}
\usepackage{rotating}
\usepackage{graphicx}
\usepackage{wrapfig}
\usepackage{braket,slashed}
\usepackage{amssymb,amsmath,bm}
\usepackage{setspace,comment,lipsum} 

\modulolinenumbers[5]

\journal{Nuclear Physics A}

\newcommand{\be}{\begin{equation}}
\newcommand{\ee}{\end{equation}}
\newcommand{\bdm}{\begin{displaymath}}
\newcommand{\edm}{\end{displaymath}}
\newcommand{\btb}{\begin{tabular}}
\newcommand{\etb}{\end{tabular}}
\newcommand{\bfig}{\begin{figure}}
\newcommand{\efig}{\end{figure}}

\newcommand{\R}{\mathrm{Re}}

\newcommand{\ddn}{D^0\bar{D}^0}
\newcommand{\ddc}{D^+D^-}
\newcommand{\dd}{D\bar{D}}
\newcommand{\dstate}{1\ ^{3}D_{1}}
\newcommand{\sstate}{2\ ^{3}S_{1}}

\begin{document}

\begin{frontmatter}

\title{Line-shape and Poles of the $\psi(3770)$}

\author[mymainaddress]{Susana Coito\corref{mycorrespondingauthor}}
\cortext[mycorrespondingauthor]{Corresponding author}
\ead{scoito@ujk.edu.pl}

\author[mymainaddress,mysecondaryaddress]{Francesco Giacosa}
\ead{fgiacosa@ujk.edu.pl}
\address[mymainaddress]{Institute of Physics, Jan Kochanowski University, Kielce, Poland}
\address[mysecondaryaddress]{Institut f\"ur Theoretische Physik, Johann Wolfgang Goethe-
Universit\"at, Frankfurt am Main, Germany}

\begin{abstract}
We study the non-Breit-Wigner line-shape of the $\psi(3770)$ resonance, predominantly a $1\ ^{3}D_{1}$ $\bar{c}c$ state, using
an unitarized effective Lagrangian approach, including the one-loop effects of
the nearby thresholds $D^+D^-$ and $D^0\bar{D}^0$. A fit of the theoretical result to the total cross-section $e^{+}%
e^{-}\rightarrow D\bar{D}$ is performed, leading to a good description of data
($\chi^{2}/d.o.f.\sim 1.03$). The partial cross sections $e^{+}e^{-}\rightarrow D^0\bar{D}^0$ 
and $e^{+}e^{-}\rightarrow D^+D^-$ turn out to be separately in good agreement with the experiment. We find a pole at $3777-i12$ MeV, that is within the Particle Data Group (PDG) mass and width estimation for this state. Quite remarkably, we find an additional, dynamically generated, companion pole at $3741-i19$ MeV, which is responsible for the deformation on the lower energy side of the line-shape. The width for the leptonic decay $\psi(3770)\rightarrow
e^{+}e^{-}$ is 112 eV, hence smaller than the PDG fit of $262\pm 18$ eV, 
yet in agreement with a recent experimental study.
 
\end{abstract}


\end{frontmatter}

\linenumbers
\section{\label{intro}Introduction}

The $\psi(3770)$ resonance is a vector state that was first detected at SPEAR \cite{prl39p526} in 1977. The signal was fitted with a pure $P$-wave
Breit-Wigner distribution. More recently, the interest on this state was
revived, and the parameters are nowadays fitted by the 
Particle Data group (PDG) as $3773.13\pm 0.35$ MeV for the mass and $27.2\pm1.0$ MeV 
for the width \cite{pdg}. According to the `quark model review' of the PDG, 
the $\psi(3770)$ resonance is classified as a $\dstate$ charmonium state \cite{prd32p189}, the first one 
above the $\dd$ threshold, which is the reason for its relatively large width. 
As a consequence, besides kinematic interference, 
additional nonperturbative effects are expected.

Indeed, the line-shape of the $\psi(3770)$ resonance turned out to be quite
anomalous. Although other experiments have been performed with observations in
$\dd$ channel
\cite{prl93p051803,prl97p121801,prd76p111105,plb668p263,plb711p292}, the
deformation on the line-shape of the $\psi(3770)$ was made clear by the BES
Collaboration data in the $e^{+}e^{-}$ annihilation to hadrons
\cite{prl101p102004}. The existence of a second resonance was suggested,
although possible dynamical effects, generated by the $\dd$ threshold were not
discarded. A deformation of a line-shape due to the superposition of two
resonances has been discussed before e.g., for the scalar kaon
\cite{npb909p418}, where, besides the dominant $\bar{q}q$ state, an additional
dynamically generated state arises from the continuum, the well known
$K_{0}^{\ast}(800)$. Such effect has not been discussed before for the vector
charmonium. 

Various analysis of the $\psi(3770)$ have been performed. In
Refs.~\cite{prd86p114013,prd87p057502} fits were computed taking into account
not only the $\dd$ interference but also the tail of the $\psi(2S)$. Such
inclusion is natural since the $\psi(3770)$ should be a mixed state $\ket{
\dstate}$-$\ket{\sstate}$. In Ref.~\cite{prd80p074001}, the
deformation from the right side of the resonance, i.e. a dip structure, is
explained by the interference with the $\dd$ kinematical background which
is higher for larger relative momentum. The same dip is reproduced in
\cite{prd86p114013}, using $\dd$ background only, in \cite{plb747p321}, 
where in addition, the continuum of light hadrons is removed, and in  Refs.~\cite{1408.5600} and \cite{1410.1375}, using Fano resonances. In \cite{plb641p145} BES measured an unexpectedly large branching fraction
$\psi(3770)\rightarrow$ non$-\dd$ of about $15\%$, a result that is not
contradicted by CLEO in \cite{prl96p092002}, within errors.\textbf{
}Predictions for such non-$\dd$ continuum were made including the tails
of $J/\psi$ and $\psi(2S)$, $\tau\tau$ and $uds$ decays \cite{plb769p187},
other excited $\psi$ states \cite{prd81p011501}, the $\dd^{\ast}$ channel
\cite{prd88p014010}, other hadronic decays and radiative decays
\cite{prd87p057502}, and final state interactions \cite{plb675p441}, which in
any case do not sum up to the value of $15\%$. Yet, since the phase space to
$\dd$ is not excessively large, it is likely that the missing decays are simply
the sum of all the many Okubo-Zweig-Iizuka (OZI)-suppressed hadronic decays, 
that have not been studied systematically in the theory. Estimations for the $\psi(3770)$
production via $p\bar{p}$ annihilation are made in
\cite{prd90p032007,epjc76p192,prd91p114022}, which may possibly be measured at
the PANDA experiment \cite{0903.3905}. Mass estimations for the $\psi(3770)$ have also been made on the lattice \cite{jhep09p089,prd77p034501}. For a review of the charmonium states, see Ref.~\cite{ppnp54p615}.

In this work, we study the properties of the $\psi(3770)$ by analyzing its
production through electron-positron annihilation, and subsequent decay into $\dd$ pairs (for a
preliminary study, see Ref.~\cite{1708.02041}). Our starting point is a
vector charmonium {\it seed} state, which gets dressed by ``clouds'' of
$\ddc$ and $\ddn$ mesons. Our aim is to study the
deformation seen on the left side of the resonance in
Ref.~\cite{prl101p102004} with mesonic loops combined with the
nearby thresholds. To this end, we use an effective relativistic Lagrangian
approach in which a single vector state $\psi\equiv\psi(3770)$ is coupled to channels $\ddc$ and
$\ddn$, as well as to lepton pairs. The propagator of $\psi$ is
calculated at the resummed one-loop level and fulfills unitarization
requirements. Then, we perform a fit of the four parameters of our approach, i.e. 
the effective couplings of $\psi$ to $\dd$ and of $\psi$ to leptons, the mass of $\psi$, 
and a cutoff responsible for the finite dimension of the $\psi$ meson, to the
experimental cross-section of the reaction $e^{+}e^{-}\rightarrow\dd$ in Refs.~\cite{plb668p263} and \cite{prl97p121801}, in the energy region up to 100 MeV above the $\ddn$ threshold. We assume that the $\psi(3770)$ resonance
dominates in this energy range. We obtain a very good description of the data, 
which in turn allows us to determine in a novel and independent way the mass, width, and branching ratios of the $\psi(3770)$. Moreover, we study in detail, to our knowledge for the first time, the poles of this state. In fact, a pole was found in Ref.~\cite{1410.1375}, using Fano resonances, at about $3778-i14$ MeV, though in lesser detail. Quite remarkably, we find \emph{two} poles for this resonance, one which roughly corresponds to the peak of the
resonance, the seed pole, and one additional dynamically generated pole,
responsible for the enhancement left from the peak, which emerges due to the
strong coupling between the seed state and the mesonic loops. This is a companion pole,
similar to the one found in the kaonic system in Ref. \cite{npb909p418}.

As a consequence of the determined parameters, we also show that the cross
sections $e^{+}e^{-}\rightarrow$ $\ddc$ and $e^{+}e^{-}\rightarrow
\ddn$ agree separately to data. Being the resonance not an ideal
Breit-Wigner one, different definitions for the partial widths are compared to
each other. As a verification of our theoretical approach, we artificially vary the
intensity of the coupling in order to show its effect on the line-shape of the
resonance. Smaller couplings lead to a narrower Breit-Wigner-like shape, while larger couplings to an even larger deformation, which eventually gives rise to
two peaks, as in Ref.~\cite{prd93p014002}. In addition, the trajectory of each pole is analyzed, confirming the effects on the line-shape.

The paper is organized as it follows. In Sec.~\ref{model} the model is introduced and
various theoretical quantities, such as the resummed propagator, spectral function, 
cross-section, and computation of the poles, are presented. In Sec.~\ref{results} we show our results:
first, we present a fit to data and examine its consequences, then we study the position and trajectory of the poles. In Sec.~\ref{conclusion}
conclusions are drawn. Some important technical details and comparative results are discussed in the Appendices.


\section{\label{model}An Effective Model}

In this section, we present the Lagrangian of the model and evaluate the most
relevant theoretical quantities, namely the resummed propagator and the spectral
functions, needed to calculate the cross section for the processes
$e^{+}e^{-}\rightarrow \ddc$ and $e^{+}e^{-}\rightarrow \ddn$, via production of the $\psi(3770)$, and with $\ddn+\ddc$ one-loops.


\subsection{\label{model.1}The strong Lagrangian and dispersion relations}

We aim to write down a Lagrangian for the strong interaction between 
a single vectorial resonance and two pseudoscalar mesons. The bare resonance 
$\psi\equiv\psi(3770)$ is identified with a bare quarkonium $c\bar{c}$ state, 
with predominant quantum numbers $N\ ^{2S+1}L_J=\dstate$, but admixtures of other 
quantum numbers, most notably the $\sstate,$ are possible. In our model, however, 
neither the quarkonium structure nor the orbital angular momentum mixing are 
explicit, since we work with mesonic degrees of freedom in the basis of total angular momentum. 
The effective strong Lagrangian density, for the charmonium state $\psi$, reads
\be
\begin{split}
\mathcal{L}_{\psi\dd}&  =ig_{\psi \ddn}\psi_{\mu
}\Big(\partial^{\mu}\ddn-\partial^{\mu}\bar{D}^{0}D^{0}%
\Big)\\
&  +ig_{\psi \ddc}\psi_{\mu}\Big(\partial^{\mu}\ddc-\partial^{\mu
}D^{-}D^{+}\Big),\label{lagi}%
\end{split}
\ee
which corresponds to the simplest interaction among $\psi$ and its main decay
products, the pseudoscalar pairs $\ddn$ and $D^{-}D^{+}$, each vertex represented in Fig.~\ref{diag2}. In this work, we shall take into account the small mass difference between
$D^{0}$ and $D^{+}$, i.e.~isospin breaking for the masses, but we shall keep the same
coupling constant, i.e.~isospin symmetry for the decays, and we define $g_{\psi D^{0}\bar{D}
^{0}}=g_{\psi \ddc}\equiv g_{\psi \dd}$. The full Lagrangian can be found in \ref{A}.

\bfig[ptb]
\centering
\btb[c]{cc}
{\small\hspace*{2.3cm}$D^0$}&{\small \hspace*{2.3cm}$D^+$}\\[0mm]
\hspace*{-1.cm}$\psi$&\hspace*{-1.cm}$\psi$\\[-3mm]
\resizebox{!}{30pt}{\includegraphics{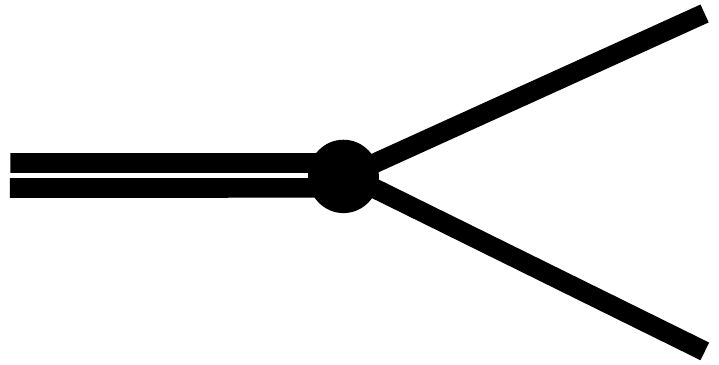}}&
\resizebox{!}{30pt}{\includegraphics{vertex1.pdf}}\\
{\small\hspace*{2.3cm}$\bar{D}^0$}&{\small \hspace*{2.3cm}$D^-$}\\[1mm]
\etb
\caption{\label{diag2}Interaction vertices $\psi\to \ddn$ and $\psi\to D^+\bar D^-$, corresponding to the Lagrangian in Eq.~\eqref{lagi}.}
\efig

\bfig[ptb]
\centering
\btb[c]{c}%
{\small\hspace*{2.3cm}$D^0$\hspace*{1.7cm}$D^+$}\\[1mm]
\resizebox{!}{18pt}{\includegraphics{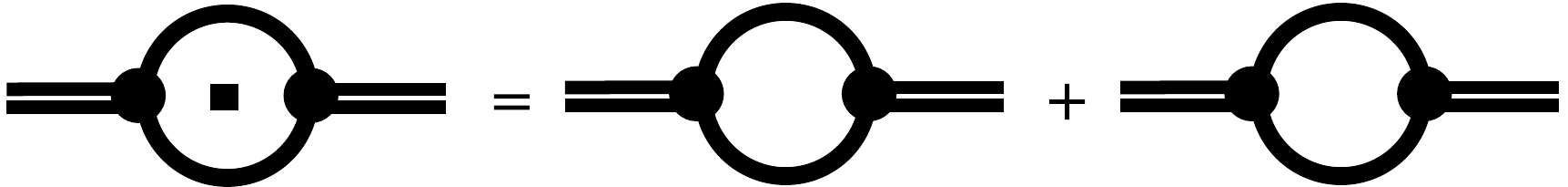}}\\[-1mm]
{\small\hspace*{2.3cm}$\bar{D}^0$\hspace*{1.7cm}$D^-$}\\[3mm]
\vspace*{2mm}
\resizebox{!}{18pt}{\includegraphics{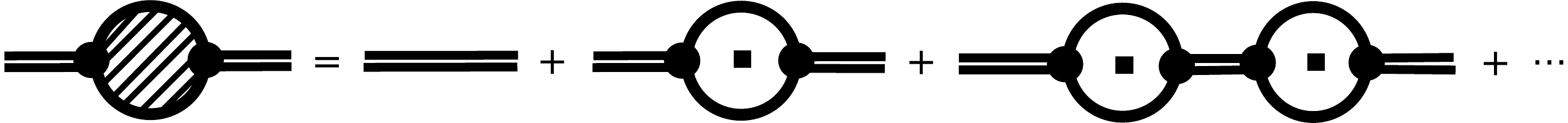}}
\etb
\caption{\label{diag3}Upper diagram: one-loop $\dd=\ddn+\ddc$.
Lower diagram: Full one-loop expansion (cf.~Fig.~\ref{diag1}).}%
\efig

The free propagator of a vector field $\psi$, without any loops, with mass $m_{\psi}$ and momentum
$p$, is given by
\be
G_{\mu\nu}(p)=\frac{1}{p^{2}-m_{\psi}^{2}+i\varepsilon}\Big(-g_{\mu\nu
}+\frac{p_{\mu}p_{\nu}}{m_{\psi}^{2}}\Big), \label{vprop}%
\ee
where the term in parenthesis is the sum over the three polarization states of
a vector. We include the resummed one-loop effect as shown in Fig.~\ref{diag3}, leading to the full propagator of $\psi$
\be
\Delta_{\mu\nu}(p)=G_{\mu\nu}(p)+G_{\mu\mu^{\prime}}%
(p)\Pi_{\mu^{\prime}\nu^{\prime}}(p)G_{\nu^{\prime}\nu}(p)+\cdots,
\label{1loop}%
\ee
where $\Pi_{\mu\nu}(p)$ is the loop-function, consisting of two
contributions, the $\ddn$ loops and the $\ddc$ loops. In
particular, we get%
\be
\Pi_{\mu\nu}(p)=g_{\psi \dd}^2\{\Sigma_{\mu\nu}(p,m_{D^{0}%
})+\Sigma_{\mu\nu}(p,m_{D^{+}})\},
\ee
where the function $\Sigma_{\mu\nu}(p,m)$, with $\int_q\equiv\int\frac{d^{4}q}{(2\pi)^{4}}$, reads
\be%
\Sigma_{\mu\nu}(p,m)=i
\int_q\frac{4\ q_{\mu}q_{\nu}f_{\Lambda}^{2}%
(\xi)}{[(q+p/2)^{2}-m^{2}+i\varepsilon][(q-p/2)^{2}-m^{2}+i\varepsilon]},
\label{se}%
\ee
where $m$ is the mass of the meson circulating in the loop (either $D^{0}$ or
$D^{+}$), $q=(q^{0},\mathbf{q})$ is the four-momentum
of the loop, and $\xi=4(\mathbf{q}^{2}+m^{2})$ (see \ref{B} for computation details). Note, in the reference
frame of the decaying particle, it holds the relation%
\be
\label{rfdp}
s=p^{2}=E^{2},\ \ p=(E,\vec{0}).%
\ee

An important element for our discussion is the vertex (or cutoff) function $f_{\Lambda}(\xi)$
entering in Eq.~\eqref{se}. This is a form-factor that is needed to
account for the unknown black vertices in Fig.~\ref{diag2}, due to the fact that
mesons are not elementary particles \cite{prc76p065204}. The cutoff function must
ensure the convergence of the integral. We choose the following Gaussian form
\be
f_{\Lambda}(\xi)=e^{-\xi/(4\Lambda^{2})}\times e^{\left(  m_{D^{0}}^{2}+m_{D^{+}%
}^{2}\right)  /\left(  2\Lambda^{2}\right)  },\label{cutoff}%
\ee
where the second term is built up for convenience. In the isospin limit,
$f_{\Lambda}$ reduces to $e^{-{\mathbf q}^{2}/\Lambda^{2}}$.
The parameter $\Lambda$, introduced in Eq.~\eqref{cutoff}, is on the same
level of all the other parameters of our model and will be evaluated through
our fit to data. Moreover, even if our form factor depends on the three-momentum
$\mathbf{q}$ only, hence strictly valid in the rest frame of the decaying
particle, covariance can be recovered by properly generalizing the vertex
function, see details in Ref.~\cite{apps9p467} 
\footnote{One can formally introduce
the vertex function already at the Lagrangian level by rendering it nonlocal, see
the discussion in Refs.~\cite{npa582p655,prd68p014011,prc71p025202,prc76p065204}, where it is also pointed out that the special form of the form factor is not important, as long as fast
convergence is guaranteed.}. However, the exponential form for the cutoff is
very typical and has been used in many different approaches in hadron physics, e.g. Ref.~\cite{prd68p014011}. For completeness, in \ref{C} we shall also present the results for a different vertex function.

We now turn back to the study of the propagator. In \ref{B}, it is shown that 
the relevant quantity is the transverse part of the propagator that, in the rest
frame of the decaying particle, and for a single loop, reads%
\begin{align}
\label{loopint}
&\Sigma(s,m)  =\frac{1}{3}\left(  -g^{\mu\nu}+\frac{p^{\mu}p^{\nu}}{p^2%
}\right)  \Sigma_{\mu\nu}(p,m)\nonumber\\
&  =-\frac{i}{3}\int_q\frac{4\ \mathbf{q}^{2}\ f_{\Lambda
}^{2}(\xi)}{[(q+\frac{p}{2})^{2}-m^{2}+i\varepsilon][(q-\frac{p}{2})^{2}-m^{2}+i\varepsilon
]}\text{,}%
\end{align}
and similarly for the sum%
\be
\Pi(s)=\frac{1}{3}\left(  -g^{\mu\nu}+\frac{p^{\mu}p^{\nu}}{p^{2}}\right)
\Pi_{\mu\nu}(p,m)\text{ }
 =g_{\psi \dd}^2\{\Sigma(s,m_{D^{0}})+\Sigma(s,m_{D^{+}})\}.
\ee
The scalar part of the one-loop resummed dressed propagator reads%
\be
\Delta(s)=\frac{1}{s-m_{\psi}^{2}+\Pi(s)}.
\ee
Note, this expression of the propagator fulfills unitarity
(cf.~Sec.~\ref{model.4} below) and it is accurate as long as further contributions to the
self-energy are small \footnote{ The next contribution would be represented by a loop in which the
unstable state $\psi$ would be exchanged by $D$ mesons circulating
in the loop. Such contributions are typically very small in hadron physics, as shown for instance in Ref.~\cite{npb888p287}. In addition, one could include also other channels, such
as many other possible but suppressed decays of the $\psi(3770)$ (e.g.,
$J/\psi\pi^{+}\pi^{-}$, etc...), as well as subthreshold channels such as
$D^{\ast}\bar{D}+h.c.$ that can contribute to the real part of the propagator.}.

The self-energy in Eq.~\eqref{se} can either be computed through the
integration (as in \ref{B}), or using dispersion relations. We follow the
latter. To this end, we first decompose $\Sigma(s,m)$ into its real and imaginary
parts
\be
\Sigma(s,m)=R(s,m)+iI(s,m),\ \ R,\ I\in\Re.\label{loop}%
\ee
According to the optical theorem, the imaginary part $I(s,m)$ (dispersive
term) is given by
\be
I(s,m)=\frac{k(s,m)}{8\pi \sqrt{s}\ g_{\psi\dd}^2}|\mathcal{M}_{\psi\rightarrow \dd}|^{2},\label{I}
\ee
with $k(s,m)=\sqrt{s/4-m^{2}}$ being the center-of-mass
momentum of the final mesons. The partial decay widths of $\psi\rightarrow
\ddn$ and $\psi\rightarrow \ddc$ are then calculated as%
\be
\Gamma_{\psi\rightarrow \ddn}(s)=\frac{k(s,m_{D^{0}})}{8\pi
s}|\mathcal{M}_{\psi\rightarrow \ddn}|^{2}\\
=g_{\psi\dd}^{2}\frac{I(s,m_{D^{0}})}{\sqrt{s}},%
\ee%
\be
\Gamma_{\psi\rightarrow \ddc}(s)=\frac{k(s,m_{D^{+}})}{8\pi
s}|\mathcal{M}_{\psi\rightarrow \ddc}|^{2}\\
=g_{\psi \dd}^{2}\frac{I(s,m_{D^{+}})}{\sqrt{s}},%
\ee
where the Lorentz invariant amplitudes squared, computed from $\mathcal{L}%
_{I}$ in Eq.~\eqref{lagi}, are given by
\begin{align}
&|\mathcal{M}_{\psi\to \ddn}|^{2}=g_{\psi\dd}^{2}\frac{4}{3}k^{2}(s,m_{D^{0}})f_{\Lambda}^{2}(s)\ ,\label{pamp1}\\
&|\mathcal{M}_{\psi\to \ddc}|^{2}=g_{\psi \dd}%
^{2}\frac{4}{3}k^{2}(s,m_{D^{+}})f_{\Lambda}^{2}(s)\text{ .}%
\label{pamp2}
\end{align}

Note, by choosing the form factor as function of the energy [$f_{\Lambda
}(\xi =4(\mathbf{q}^{2}+m^{2}))$], when the imaginary part of the loop in Eq.~\eqref{loopint} is
taken, the replacement $\xi \rightarrow s$ is performed (see Eqs.~\eqref{pamp1} and \eqref{pamp2}). Then,
the function $f_{\Lambda }(s)$ directly  enters in various expressions.

Finally, the on-shell partial and total decay widths are
\begin{align}
\label{osw1}
&\Gamma_{\psi\rightarrow \ddn}^{\text{on-shell}}=\Gamma
_{\psi\rightarrow \ddn}(m_{\psi}^{2})\ ,\\
\label{osw2}
&\Gamma_{\psi\rightarrow \ddc}^{\text{on-shell}}=\Gamma_{\psi\rightarrow
\ddc}(m_{\psi}^{2})\ , \\
\label{oswt}
&\Gamma_{\psi\to \dd}^{\text{on-shell}}=\Gamma_{\psi\rightarrow \ddn}^{\text{on-shell}}+\Gamma_{\psi\rightarrow \ddc}^{\text{on-shell}}\text{
}.
\end{align}
Once the imaginary part of the loop is known, the real part, the function
$R(s,m)$, is computed from the dispersion relation
\be
R(s,m)=\frac{PP}{\pi}\int_{s_{th}=4m^2}^{\infty}\frac{I(s^{\prime},m)}%
{s^{\prime}-s}\ \mathrm{d}s^{\prime},\label{dr}%
\ee
where $I(s^{\prime},m)$ is zero below threshold. Convergence is
guaranteed by the cutoff function. As we shall see in Sec.~\ref{results}, the real part $R(s,m)$ causes a distortion
in the line-shape due to the continuous shifting of the physical mass of the
resonance with the energy. Note, when the energy $E=z$ is complex, the function 
$\Sigma(s=z^{2},m)$ reads (away from the real axis):%
\be
\Sigma(z^{2},m)=\frac{1}{\pi}\int_{s_{th}=4m^2}^{\infty}\frac{I(s^{\prime}%
,m)}{s^{\prime}-z^{2}}\ \mathrm{d}s^{\prime}.
\label{eq21}
\ee

This formula clearly shows that in the first Riemann
Sheet (RS) the complex function $\Sigma(z^{2},m)$ is regular everywhere on the $z^{2}$-complex plane, besides a cut from $4m^{2}$
to infinity. In particular,
$\Sigma(z^{2}\rightarrow\infty,m)\to 0$ in all directions \footnote{In order to avoid misunderstanding, we recall that, in the 1st Riemann Sheet, the function $\Sigma
(z^{2},m)$ is an utterly different complex function than $f_{\Lambda}^{2}(z^{2})\propto
e^{-z^{2}/2\Lambda^{2}}$. Namely, Eq.~\eqref{eq21} implies
solely that $\operatorname{Im}\Sigma(x^{2},m)$ contains $e^{-x^{2}/2\Lambda^{2}}$ for $z^{2}=x^{2}$ being real. This fact is also
clear by noticing that, while $\Sigma(z^{2}\rightarrow\infty,m)\to 0$ in any
direction, $f_{\Lambda}^{2}(z^{2}\rightarrow\infty)$ diverges for
$\operatorname{Re}\ z^{2}<0$.}.
Furthermore, although it is not strictly needed, since the integral in
Eq.~\eqref{dr} is already convergent, we use the once-subtracted dispersion
relation, with subtraction in point $m_{\psi}^{2}$, for our convenience.
Hence, the total loop-function is given by
\be
\Pi_1(s)=\Pi(s)-R(m_{\psi}^{2},m_{D^{0}})-R(m_{\psi}%
^{2},m_{D^+})\text{ ,}\label{1sub1}%
\ee
and, the final dressed propagator of the $\psi$ meson is%
\be
\Delta(s)=\frac{1}{s-m_{\psi}^{2}+\Pi_1(s)}\text{ }%
.\label{propfinal}%
\ee
In this way, the parameter $m_{\psi}$ in the propagator corresponds to the mass
of the particle defined as%
\be
\operatorname{Re}\ \Delta_{\psi}(s)^{-1}=0\rightarrow s=m_{\psi}^{2}%
\text{.}\label{mass1}%
\ee
Other definitions for the mass are possible, such as the position of the peak, or the real part of the pole, as we shall see below.


\subsection{\label{model.2} Poles}

In order to find poles, the energy $E$ is analytically continued to the complex plane ($E\rightarrow z$). In case of two decay channels, the Riemann surface is composed by four RSs. Poles are found in the unphysical sheet which results in the physical sheet when the energy is real and the resonance can be seen. Above both thresholds, this corresponds to the condition $\mathrm{Im}\ k_{D^0}<0$ and $\mathrm{Im}\ k_{D^+}<0$ when $\R\ E>2m_{D^0}$ and $\R\ E>2m_{D^+}$, i.e., the third RS. Poles are given when the denominator of the propagator in Eq.~\eqref{propfinal} is zero in the correct RS, i.e.,
\be
z^{2}-m_{\psi}^{2}+\Pi_{1, III}(z^2)=0,\ z\in\mathbb{C},
\ee
where
\be
\Pi_{1, III}(z^2)=\Pi_{1,I}(z^{2})
-2ig_{\psi \dd}^2[I_{I}(z^{2},m_{D^{0}})+I_{I}(z^{2},m_{D^{+}})]=0\ ,
\ee
where the subscripts $I$ and $III$ stand for the first and third RS \footnote{ In other terms, $z_{pole}$ corresponds to $\sqrt{s_{pole}},$ see Eq.~\eqref{rfdp}.}. While in the first RS $\Pi_{1}(z^{2})$ is regular everywhere apart
from the cut(s) on the real axis (see Eq.~\eqref{eq21} and subsequent discussion), in
the other RSs the imaginary part continued to complex plane,
$I(z^{2},m),$ appears. As a consequence, the vertex function $f_{\Lambda}%
^{2}(z^{2})\propto e^{-z^{2}/2\Lambda^{2}}$ generates a singular point at the
complex infinity. When other choices for the vertex function are made, such as the one in \ref{C},
other singularity types are generated. In general, any nontrivial function
$f_{\Lambda}^{2}(z^{2})$ will display singularities appearing in RSs different from the first one.


\subsection{\label{model.3}Coupling to leptons}

The available experimental data comes from the production process $e^{+}e^{-}\rightarrow\psi\rightarrow
\dd$ (see Fig.~\ref{diag1}), therefore we need to couple the state $\psi(3770)$ to leptons. The
corresponding interaction Lagrangian is defined by
\be
\mathcal{L}_{\psi l^{+}l^{-}}=g_{\psi e^{+}e^{-}}\ \psi_{\mu}\sum_{l=e%
,\mu,\tau}\bar{\Psi}_{l}\gamma^{\mu}\Psi_{l},\label{lagii}%
\ee
which is the simplest interaction among a massive vector field $\psi$ and a
fermion ($\Psi$)-antifermion pair. The coupling $g_{\psi e^{+}e^{-}}$, between $\psi$ and the electron-positron pair, is here considered to be the same between $\psi$ and all leptonic pairs. It is the overall strength for the annihilation of the leptonic pair $l^+l^-$ into one photon, and further conversion into the $\psi$ vector. In Eq.~\eqref{lagii}, we describe the process $e^{+}e^{-}%
\rightarrow\psi$ through a single effective vertex proportional to $g_{\psi
e^{+}e^{-}}$, but, as shown in Fig.~\ref{diag4}, this coupling constant emerges via an
intermediate virtual photon that converts into a charmonium
state.

\bfig[ptb]
\centering
\btb
[c]{c}%
{\small\hspace*{-4cm}$e^+$\hspace{3.2cm}$e^+$}\\[2mm]
\hspace*{.9cm}$\psi$\hspace*{3.2cm}$\gamma$\hspace*{.9cm}$\psi$\\[-6mm]
\hspace*{-4mm}\resizebox{!}{55pt}{\includegraphics{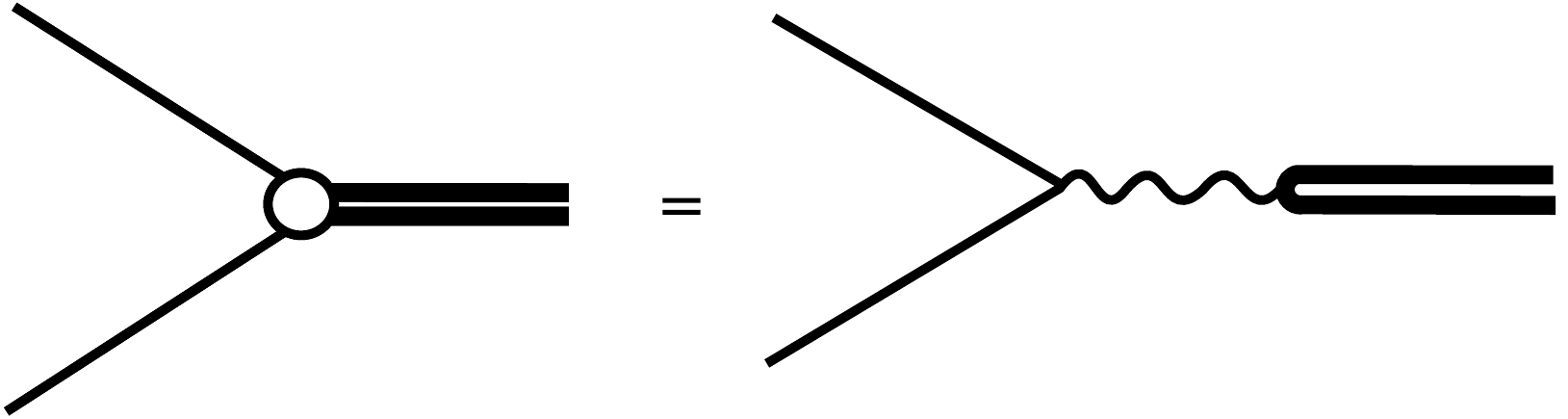}}\\[-1mm]
{\small\hspace*{-4cm}$e^-$\hspace{3.2cm}$e^-$}\\[1mm]
\etb
\caption{\label{diag4} Interaction vertex $e^{-}e^{+}\to\psi(3770)$.}%
\label{diag4}%
\efig

\bfig[ptb]
\centering
\btb[c]{c}%
{\small\hspace*{0cm}$e^-$\hspace*{3.5cm}$D$}\\
\resizebox{!}{40pt}{\includegraphics{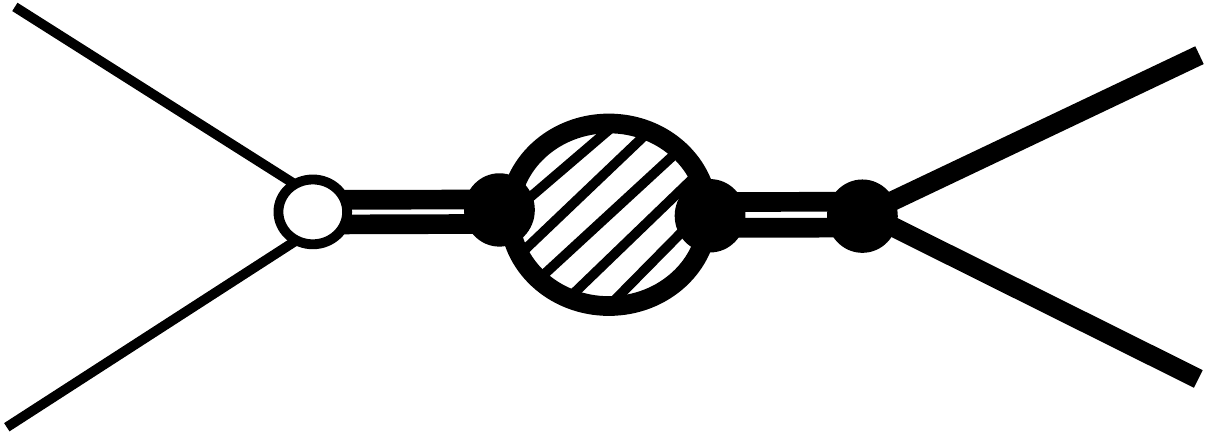}}\\[-1mm]%
{\small\hspace*{0cm}$e^+$\hspace*{3.5cm}$\bar{D}$}\\[1mm]
\etb
\caption{\label{diag1}Reaction $e^{-}e^{+}\rightarrow\psi(3770)\rightarrow \dd$ with $\ddn+\ddc$ loops (cf.~Figs.~\ref{diag2}, \ref{diag3} and \ref{diag4}).}%
\efig

The corresponding decay into leptons reads
\be
\Gamma_{\psi\rightarrow l^{+}l^{-}}(m_\psi^2)=\frac{k(m_\psi^2,m_l)}{8\pi m_\psi^2}\frac{4}%
{3}\Big(m_\psi^2+2m_{l}^{2}\Big)g_{\psi e^{+}e^{-}}^{2}, \label{gamll}%
\ee
where $m_{l}$ is the leptonic mass. In principle, the loops of leptons should be included in the total
one-loop function of resonance $\psi(3770)$, obtaining $\Pi_{tot}%
(s)=\Pi(s)+\sum_{l}\Pi_{l}(s)+\cdots$, where dots refer to other possible but suppressed hadronic decay channels. However, the loop contribution of $\Pi_{l}(s)$ is definitely
negligible w.r.t. $\Pi(s).$ Here, we shall simply consider $\Pi_{tot}%
(s)=\Pi(s).$


\subsection{\label{model.4}Spectral function and cross section}

The unitarized spectral function as a function of energy $E$, equivalent to the running mass of the $\psi(3770)$ resonance, is given by
\be
\begin{split}
&d_{\psi}(E) =-\frac{2E}{\pi}\mathrm{Im}\ \Delta(E)\\
&  =\frac{2E^{2}}{\pi}\frac{\Gamma_{\psi\rightarrow \ddn}%
(E^{2})+\Gamma_{\psi\rightarrow \ddc}(E^{2})}{[E^{2}-m_{\psi}%
^{2}+\operatorname{Re}\Pi_{1}(E^{2})]^{2}+[\operatorname{Im}\Pi
_{1}(E^{2})]^2}\text{ },
\end{split}
\ee
that has the general shape of a relativistic Breit-Wigner distribution,
distorted by the loop-function $\Pi_1$ (cf.~Eq.~\eqref{1sub1}). When no poles below threshold emerge, the
normalization above threshold is guaranteed (unitarity):%
\be
\label{unit}
\int_{2m_{D^{0}}}^{\infty}d_{\psi}(E)dE=1\text{ .}%
\ee
The quantity $d_{\psi}(E)dE$ is interpreted as the probability that the state
$\psi(3770)$ has a mass between $E$ and $E+dE.$ The cross section for $e^{+}e^{-}\rightarrow \dd=\ddc+\ddn$ takes the form
\be
\sigma_{e^{+}e^{-}\rightarrow \dd}=\frac{\pi}{2E}g_{\psi e^{+}e^{-}}%
^{2}d_{\psi}(E)=-g_{\psi e^{+}e^{-}}^{2}\mathrm{Im}\Delta(E)\text{
.}\label{cs}%
\ee
Hence, the experimental data for this cross section give us direct access to
the imaginary part of the propagator of the meson $\psi(3770)$. The
corresponding amplitude, leading to $\sigma_{e^{+}e^{-}\rightarrow \dd}$,
is depicted in Fig.~\ref{diag1}.

One also defines the partial spectral functions as \cite{fp42p1262}:%
\begin{align}
&d_{\psi\rightarrow \ddc}(E)=\frac{2E^{2}}{\pi}\frac{\Gamma_{\psi\rightarrow \ddc}(E^{2})}
{[E^{2}-m_{\psi}^{2}+\operatorname{Re}\Pi_{1}(E^{2})]^{2}+[\operatorname{Im}\Pi_{1}(E^{2}%
)]^2}\ ,\\
&d_{\psi\rightarrow \ddn}(E)=\frac{2E^{2}}{\pi}
\frac{\Gamma_{\psi\rightarrow \ddn}(E^{2})}{[E^{2}-m_{\psi}%
^{2}+\operatorname{Re}\Pi_{1}(E^{2})]^{2}+[\operatorname{Im}\Pi
_{1}(E^{2})]^2}\ .%
\end{align}
Then, the partial cross sections are given by:%

\begin{align}
&\sigma_{e^{+}e^{-}\rightarrow \ddc}  =\frac{\pi}{2E}g_{\psi
e^{+}e^{-}}^{2}d_{\psi\rightarrow \ddc}(E)\text{ ,}\\
&\sigma_{e^{+}e^{-}\rightarrow \ddn}  =\frac{\pi}{2E}g_{\psi
e^{+}e^{-}}^{2}d_{\psi\rightarrow \ddn}(E)\text{ ,}\\
&\sigma_{e^{+}e^{-}\rightarrow \dd}=\sigma
_{e^{+}e^{-}\rightarrow \ddc}+\sigma_{e^{+}e^{-}\rightarrow D^{0}\bar
{D}^{0}}\ .
\end{align}



\section{\label{results} Line-shapes and poles}

In this section, we present our fit of Eq.~\eqref{cs} to experimental data and its consequences, and the notable existence, position, and trajectory of two poles underlying the $\psi(3770)$. 


\subsection{Fit to data and consequences}

As a first necessary step, we determine the four free parameters of our model%
\be
\label{para}
\{g_{\psi\dd},\text{ }m_{\psi},\text{
}\Lambda,\text{ }g_{\psi e^{+}e^{-}}\}\text{ ,}%
\ee
defined in Sec.~\ref{model},
by performing a fit to the cross section data for the process $e^{+}e^{-}%
\rightarrow\psi\rightarrow \dd=\ddn+\ddc$. We use 14 experimental points published in Ref.~\cite{plb668p263}, and the theoretical expression in Eq.~\eqref{cs}. Note, we use only the data of Ref.~\cite{plb668p263} for the fit, since the data of Ref.~\cite{prl97p121801} are contained in the data samples of Ref.~\cite{plb668p263} (the data sets of March 2001 quoted in
these two papers are the same \cite{pc}). The fit to data is shown in Fig.~\ref{fitsum}, and the values of the
parameters in the set \eqref{para} are presented in Table \ref{fitpar}. We get the value $\chi^{2}/d.o.f.\simeq 1.03$, which shows that a very good description of the data is achieved. 

The errors of the parameters entering the fit are estimated by the square root of the corresponding Hessian matrix (this is a result of the standard procedure according to in which one defines a new set of
parameters that diagonalize the Hessian matrix, see e.g.~Ref.~\cite{prd96p054033} for
details).

There are various consequences of the fit that we discuss in detail:\\

\bfig[ptb]
\begin{center}%
\btb
[c]{c}%
\resizebox{!}{200pt}{\includegraphics{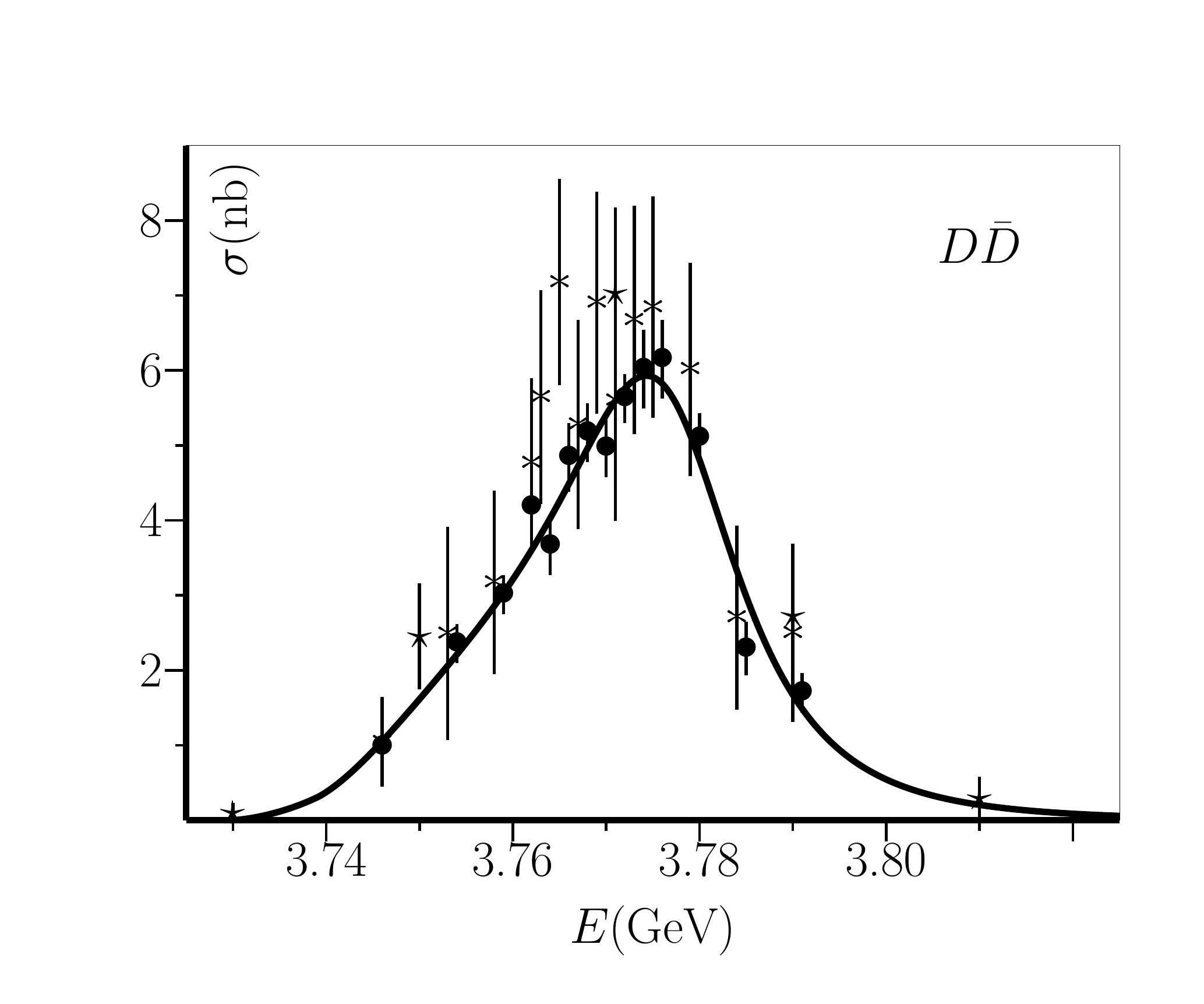}}
\etb
\end{center}
\caption{Data: $\bullet$ BES \cite{plb668p263}, $\ast$ BES \cite{prl97p121801}%
, $\star$ BaBar \cite{prd76p111105}. Solid line: our fit to data in Ref.~\cite{plb668p263} (cf.~Table \ref{fitpar}).}%
\label{fitsum}%
\efig

\begin{table}
\begin{center}
\btb{c c}
\hline
$m_\psi$ (MeV)&$3773.19\pm 0.46$\\[1mm]
$\Lambda$ (MeV)&$272.3\pm 2.3$\\[1mm]
$g_{\psi\dd}$&$30.73\pm 0.58$\\[1mm]
$g_{\psi e^+e^-}$&$(1.055\pm 0.015)\times 10^{-3}$\\[1mm]
$\chi^2$&$10.28$\\[1mm]
$\chi^2/d.o.f$&$1.028$\\
\hline
\etb
\caption{\label{fitpar} Fitting parameters.}
\end{center}
\end{table}

\bfig[ptb]
\begin{center}%
\btb
[c]{c}%
\resizebox{!}{200pt}{\includegraphics{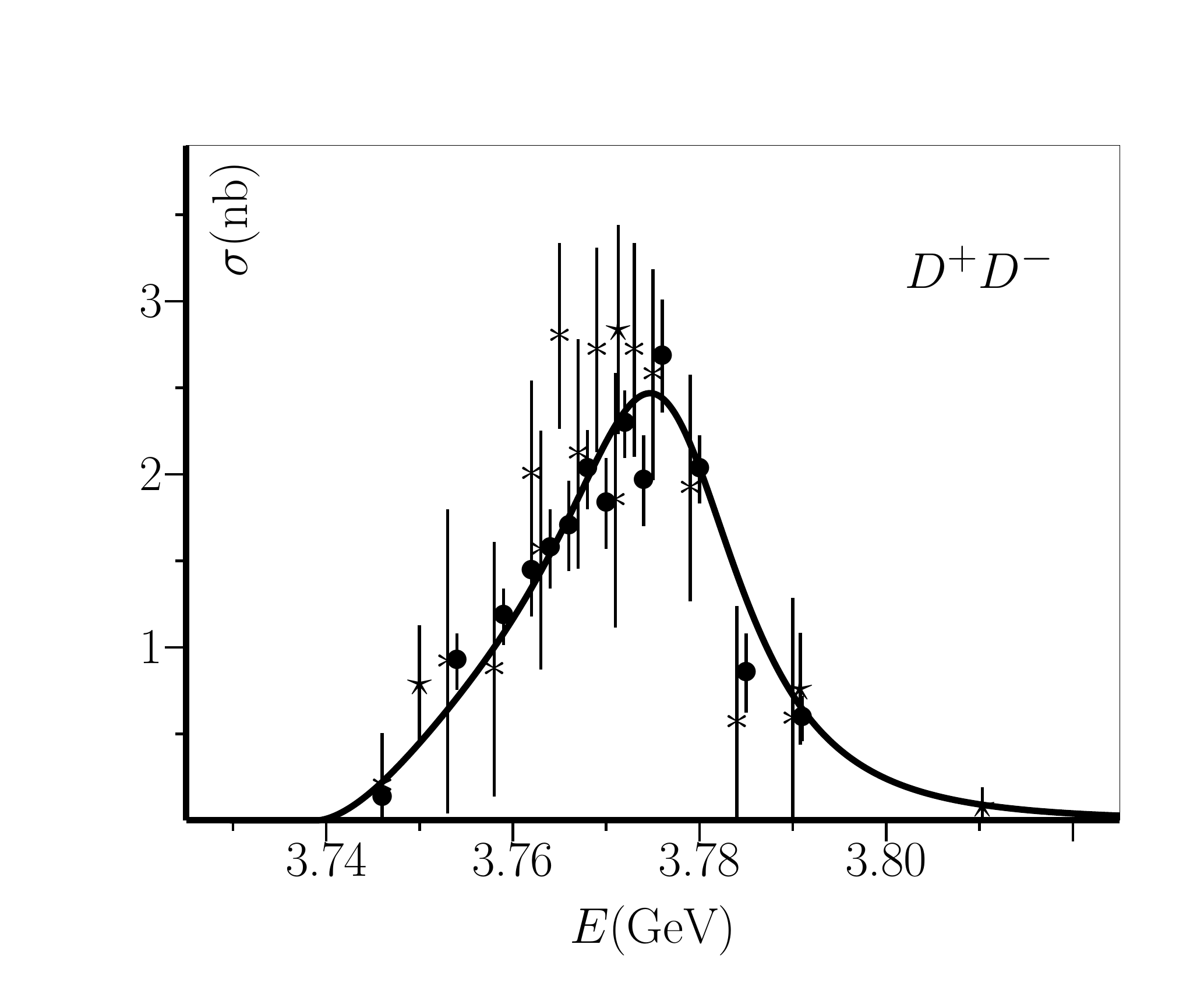}}\\[-10mm]%
\resizebox{!}{200pt}{\includegraphics{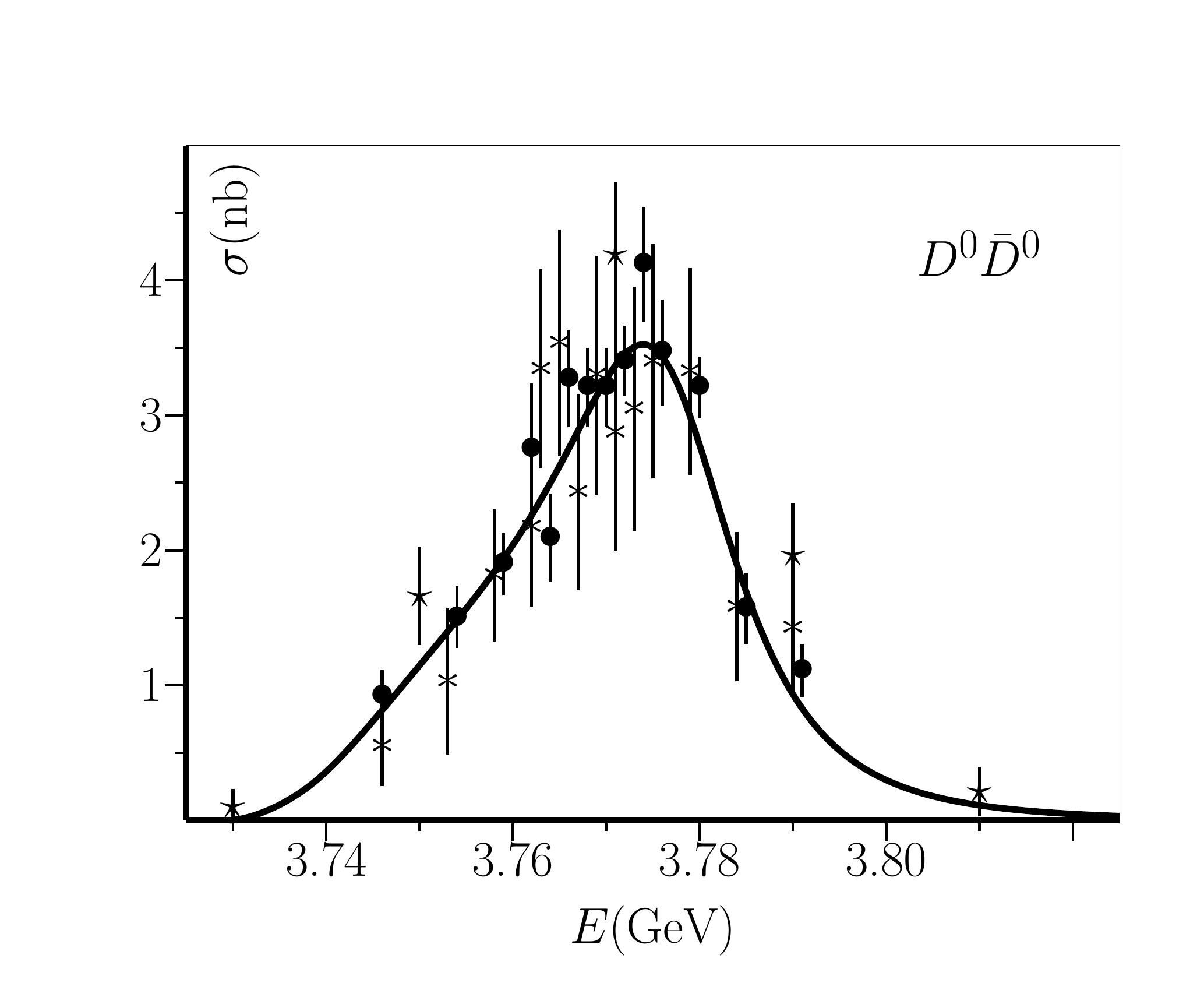}}\\[0mm]%
\etb
\end{center}
\caption{Data: $\bullet$ BES \cite{plb668p263}, $\ast$ BES \cite{prl97p121801}%
, $\star$ BaBar \cite{prd76p111105}. Solid line: our result, using the parameters in Table \ref{fitpar}.}%
\label{fit}%
\efig

\begin{table}
\begin{center}
\btb{ccc}
\hline
&$\Gamma(m_\psi^2)$ &$\Gamma^{\mathrm{average}}$\\[1mm]
\hline
$\Gamma_{\ddn}$ (MeV)&$11.5\pm 0.4$&$10.87$\\[1mm]
$\Gamma_{\ddc}$ (MeV)&$8.0\pm 0.3$&$6.95$\\[1mm]
$\Gamma_{\dd}$ (MeV)&$19.5\pm 0.7$&$17.83$\\[1mm]
$\Gamma_{\ddn}/\Gamma_{\ddc}$&$1.44$&$1.57$\\[1mm]
$\Gamma_{\ddn}/\Gamma_{\dd}$&$0.59$&$0.61$\\[1mm]
$\Gamma_{\ddc}/\Gamma_{\dd}$&$0.41$&$0.39$\\[1mm]
\hline
\etb
\caption{\label{widths} Decay widths and branching ratios using Eqs.~\eqref{osw1}-\eqref{oswt}, for the on-shell widths, and Eqs.~\eqref{avw1}-\eqref{avw3}, for the average widths.}
\end{center}
\end{table}

a) The value of the mass $m_{\psi}=$ $3773.19\pm 0.46$ MeV (corresponding to the
zero of the real part of the inverse's propagator, see Eq. (\ref{mass1})) is
in well agreement with the current PDG fit of $3773.13\pm0.35$ MeV
\cite{pdg}.

\bigskip

b) In Fig.~\ref{fit} we present a comparison between theory and the data for the cross
sections $\sigma_{e^{+}e^{-}\rightarrow \ddc}$ and $\sigma_{e^{+}%
e^{-}\rightarrow \ddn}$, using the parameters in Table \ref{fitpar}. The good agreement shows that the theory
is able to describe the two partial cross sections separately, without the need
of some extra parameter, and in agreement with isospin symmetry.

\bigskip

c) The branching ratios and partial and total ($\dd$) widths are presented in Table \ref{widths}. The partial widths, evaluated on-shell, are defined in Eqs.~\eqref{osw1}-\eqref{oswt}. The branching ratios are in agreement
with those quoted in PDG for the $\dd$. The
ratio $\Gamma_{\psi\rightarrow \ddn}^{\text{on-shell}}/\Gamma_{\psi\rightarrow \ddc}^{\text{on-shell}}$ of $1.44$ is
compatible with the PDG values, which range from $1.04$ to $1.51$. However, as
shown in Fig.~\ref{gam}, the functions $\Gamma_{\psi\rightarrow \ddn%
}(E^2)$ and $\Gamma_{\psi\rightarrow \ddc}(E^2)$ strongly vary in the region
of interest. Alternatively, one may use a different definition, cf.~\cite{prc76p065204}, to obtain the partial
decay widths and the full decay width to $\dd$, by integrating over the spectral function as
\begin{align}
&\Gamma_{\psi\rightarrow \ddn}^{\text{average}}=\int
_{2m_{D^{0}}}^{\infty}\Gamma_{\psi\rightarrow \ddn}(E^{2})\ d_{\psi
}(E)\ dE\ ,\label{avw1}\\
&\Gamma_{\psi\rightarrow \ddc}^{\text{average}}=\int_{2m_{D^{+}}%
}^{\infty}\Gamma_{\psi\rightarrow \ddc}(E^{2})\ d_{\psi}(E)\ dE\ ,\label{avw2}\\%
&\Gamma_{\psi\to\dd}^{\text{average}}=\Gamma_{\psi\rightarrow \ddn%
}^{\text{average}}+\Gamma_{\psi\rightarrow \ddc}^{\text{average}}\label{avw3}\ .
\end{align}

\bfig[ptb]
\begin{center}
\resizebox{!}{200pt}{\includegraphics{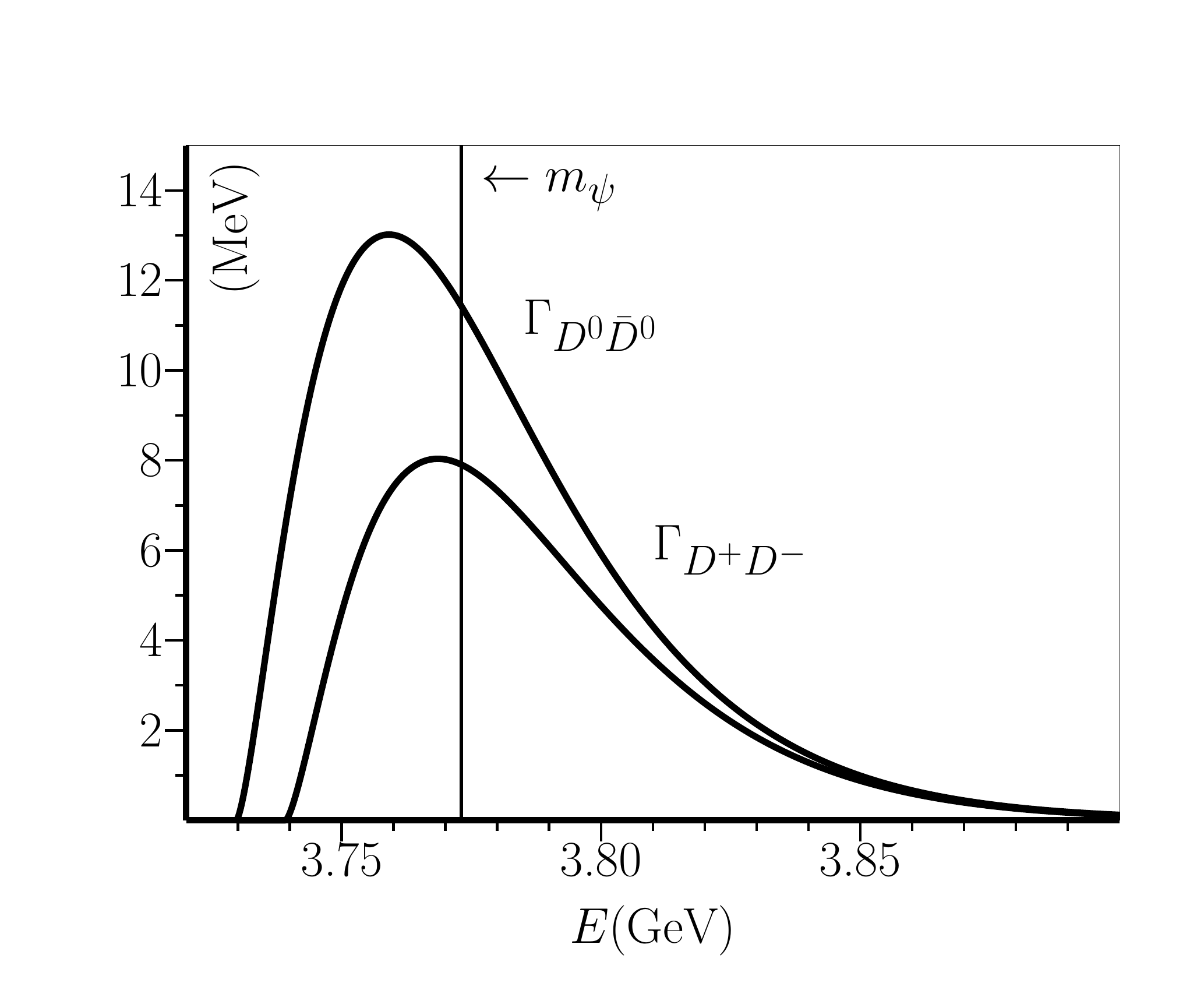}}\\[5mm]
\end{center}
\caption{Partial widths using the cutoff function in Eq.~\eqref{cutoff}, using parameters in Table \ref{fitpar}. The width $\Gamma_{\dd}$ is the sum of the width in each channel.}%
\label{gam}%
\efig

A somewhat naive but still useful alternative determination of the mass is the
value corresponding to the peak and the half-height width:%
\be
\begin{split}
&m_{\psi}^{\text{peak}}\simeq 3774.3\pm 1.0\text{ MeV, }\\
&\Gamma_{\psi\to \dd}^{\text{half-height}}\simeq 21.5\pm 1.0\text{ MeV ,}
\end{split}
\ee
where we estimated the error of the peak mass as $\sim 1$
MeV, close to the determined error for the parameter $m_{\psi}$, and the error
of the distribution width as $\sim 1$ MeV as well, coincident to the error for
$\Gamma_{D\bar{D}}$ (see Table \ref{widths}).
These evaluations show that the resonance $\psi(3770)$ is far from being an ideal Breit-Wigner.
While all these different approaches coincide for states with a very small
width, here the distortions are sizable, hence clear definitions of mass and width, 
as well as branching ratios, are difficult. A commonly used approach to get
a uniform result, makes use of poles in the complex plane, as we show in Sec.~\ref{results.2}.

This discussion shows that there is no unique definition
of the total and the partial widths. This fact also renders a direct
comparison with the fit of the PDG quite difficult. Namely, according to
PDG, one has $\Gamma _{\psi (3770)\rightarrow D^{0}\bar{D}%
^{0}}^{PDG}=14.1_{-1.5}^{+1.2}$ MeV, $\Gamma _{\psi (3770)\rightarrow
D^{+}D^{-}}^{PDG}=11.2_{-1.7}^{+1.7}$ MeV, and $\Gamma _{\psi (3770)\rightarrow
DD}^{PDG}=25.2_{-2.6}^{+2.1}$ MeV \cite{pdg}, hence our results for the partial decay
widths evaluated on shell and presented in Table \ref{widths}, first column, are
somewhat smaller. However,
as visible in Fig.~\ref{gam}, the corresponding function vary strongly in the energy
region of interest. Moreover, instead of using theoretical partial widths
which cannot be uniquely defined, one should stress that our approach
correctly describes the data for channels $D^{0}\bar{D}^{0}$ and $%
D^{+}D^{-}$ separately, as visible in Fig.~\ref{fit}, and as described in point b)
above.

Another commonly used approach to circumvent all these definition problems
mentioned above, is to move away from the real axis and to study the pole(s)
in the complex plane. As we shall see in detail in Sec.~\ref{results.2}, the seed
pole of $\psi (3770)$ corresponds to a larger decay width of $24.6$ MeV
which is in very well agreement with PDG.

\bigskip

d) Using the coupling $g_{\psi e^{+}e^{-}}$ that outcomes from the fit, the
widths $\Gamma_{\psi\rightarrow l^{+}l^{-}}$ can be easily computed from
Eq.~\eqref{gamll}. Results are shown in Table \ref{lw}. The value for
$\Gamma_{\psi\rightarrow e^{+}e^{-}}$ is smaller than the one given by the PDG fit of $262\pm 18$ eV  
by a factor two, but it is compatible with the analysis in Ref.~\cite{plb711p292}, that gives $154^{+79+21}_{-58-27}$ eV. The mismatch of our result with the PDG estimate, that nevertheless lists a quite broad range of values from different experiments, shows that this decay rate should be further
investigated in the future. Also the experimental identification of the decays
into $\mu^{+}\mu^{-}$ and $\tau^{+}\tau^{-}$ could help to distinguish among
different models.

\begin{table}
\centering
\btb{ccc}
\hline
$\Gamma_{e^+e^-}$ (eV)&$\Gamma_{\mu^+\mu^-}$ (eV)&$\Gamma_{\tau^+\tau^-}$ (eV)\\[1mm]
\hline
$111.4\pm 3.3$&$111.4\pm 3.3$&$54.0\pm $\\[1mm]
\hline
\etb
\caption{\label{lw}Leptonic widths computed with Eq.~\eqref{gamll}.}
\end{table}

\bigskip

e) The Gaussian vertex function together with the cutoff
value $\Lambda =272.3\pm 2.3$ MeV takes into account in a simple way the
composite nature of the resonance $\psi(3770)$ and its nonlocal interaction
with the $D$ mesons. Microscopically the vertex function emerges from the
nonlocal nature of both $\psi (3770)$ and of the decay products $D$ (for
point-like part interaction, $\Lambda $ goes to infinity), see Refs.~\cite{prd68p014011,0903.3905}
for an explicit treatment. (Indeed, the rather small value of $\Lambda $
emerging from our fit is intuitively understandable by the fact that the
resonance $\psi (3770)$ is predominantly a $D$ wave. A detailed study of
this issue by using a microscopic model such as the one in Ref.~\cite{prd94p014016} will be
given elsewhere). The vertex function is a crucial ingredient that defines our
model and, among other properties, guarantees the finiteness of the results.
The exponential form used here emerges naturally from various microscopic
approaches, see Refs.~\cite{prd94p014016,prd53p295} and refs.~therein. However, our results do not
strongly depend on the precise choice of the vertex function as long as it is smooth and at the same time 
falls sufficiently fast. As we show in \ref{C}, a hard cutoff would imply that the spectral function would fall abruptly to zero above a certain threshold; this unphysical behavior does not lead to any satisfactory description of data. On the other hand, as
discussed in \ref{D}, the avoidance of a form factor through a three-times
subtracted dispersion relation does not lead to satisfactory results, thus
confirming that the need of some vertex function for our treatment of
mesonic loops is needed. Once the vertex function is fixed, the model is
mathematically consistent at any energy: in fact, the important
normalization condition reported in Eq.~\eqref{unit} is obtained by formally
integrating up to infinity (numerically, the normalization is verified by
integrating up to $10$ GeV). In turn, this means that, from a mathematical
perspective, the value of the momentum of the emitted mesons $D$ can be
larger than $\Lambda .$ A different issue is the maximal energy up to which
we shall trust our model. In fact, even if mathematically consistent, the
model is limited because it takes into account only one vector resonance and
neglects the contributions of other resonances, most notably $\psi (2S),$ as
well as the background (see the comments h) and i) about these topics) and
the next opening thresholds, such as $D\bar{D}^{\ast }$. We therefore trust the
model in the energy range starting from the lowest threshold, $2m_{D^{0}}$, 
up to at most $3.8$ GeV, when the line-shape goes down and, as described
in Refs.~\cite{prd86p114013,prd87p057502}, the role of the background becomes
important.

\bigskip

f) In our fit, we consider only the $\psi(3770)$ resonance as a virtual state
of the process $e^{+}e^{-}\rightarrow \dd$. As discussed above, a possible mixing between bare
$c\bar{c}$ states with quantum numbers $\dstate$ and $\sstate$ is automatically taken into account in our bare field $\psi$
entering into the Lagrangian defined in Eq.~\eqref{lagi}. Indeed, the bare field, i.e., the field prior
to the dressing by mesonic loops, represents the diagonal quark-antiquark state, while the $\psi(3770)$ is a mixture of $\dstate$ and $\sstate$ configurations, with the $\psi(2S)$ as its orthogonal state. Yet, the role of the $\psi(2S)$ as an additional exchange in the reaction
$e^{+}e^{-}\rightarrow\psi(2S)\rightarrow \dd$ was not taken into account.
Being $\psi(2S)$ off-shell in the energy of interest, the propagator is
expected to suppress this amplitude; moreover, also its coupling to channel $\dd$
is expected to be small, due to the form-factor. Indeed, in
Refs.~\cite{plb718p1369,prd81p034011}, the $\psi(2S)$, $\dd$ loops, and
$D\bar{D}$ rescattering have been considered, leading to fits that are
in good agreement with the experiment in Ref.~\cite{plb668p263}. In both references the $\psi(2S)$ contribution is quite small, in agreement with our
results. However in their case the $\psi(2S)$ was still necessary to obtain a good fit, whereas in our model it is not. 
Note, in Refs.~\cite{plb718p1369,prd81p034011} the pole structure was not examined.

\bigskip

{g) In this work we did not consider a four-leg interaction process of
the type $D\bar{D}\rightarrow D\bar{D},$ that can be formally described by a
four-vertex proportional to $(\partial_{\mu}D\bar{D}-\partial_{\mu}\bar
{D}D)^{2}$. Such a rescattering describes elastic scattering, induced by a direct four-body interaction, and also as an effective description of various $t$-channel exchanges between $D\bar{D}$. Our model can
describe properly the data without this contribution, hence the inclusion of a four-leg interaction is not needed to
improve the agreement with the data. This does not necessarily mean that its
magnitude is small, but simply that one can hardly disentangle its role from
the $D\bar{D}$-loops that we have considered in our approach. In addition,
there is a more formal point that should be taken into account: one can in
principle redefine the field $\psi_{\mu}\rightarrow\psi_{\mu
}+\alpha(\partial_{\mu}D\bar{D}-\partial_{\mu}\bar{D}D),$ where $\alpha$ is a
parameter on which no physical quantity should depend on \cite{pr130p776,epja44p93}. The
redefinition would however lead to some rescattering terms. Of course, the
independence on the field redefinition can be fully accomplished only if one
could solve the theory exactly, since in any approximation scheme differences
would still persist at a given order. In conclusion, the study of 
a four-leg interaction term (as well as other effects such as the mixing with $\psi(2S)$) should be performed when more precise data will be available.

\bigskip

h) Here, we discuss more closely the
differences between our work and the one in Ref.~\cite{plb718p1369}. The main difference is the
treatment of the propagator of the $\psi(3770)$ meson, in particular for what
concerns the self-energy. In Ref.~\cite{plb718p1369} the imaginary part of the self-energy
(i.e., the energy-dependent decay width) grows indefinitely with increasing
$s.$ In this way, the K\"{a}ll\'{e}n--Lehmann representation does not lead to
a normalized spectral function. Moreover, the real part of the self-energy
depends on an additional subtraction constant $\mu$ (linked to the
regularization of the loop in Ref.~\cite{plb718p1369}, hence ultimately corresponding to
our cutoff $\Lambda$).\ When $\mu$ is set to zero (their so-called minimal
subtraction scheme) the real part of the self energy simply vanishes (hence,
it is trivial). When $\mu$ is different from zero, the real part also grows
indefinitely for increasing $s.$ Indeed, in the approach of Ref.~\cite{plb718p1369}, the
propagator of $\psi(3770)$ is not enough to describe data, also when the
rescattering is present; the inclusion of the $\psi(2S)$ state is necessary.
Hence, our approach is different under many aspects: since it contains a fully
consistent treatment of $D\bar{D}$ loops, the dispersion relations are
fulfilled and the normalization of the spectral function is naturally
obtained. In this way, nontrivial effects which modify the form of the
spectral function automatically arise. Moreover, in our case neither the $\psi(2S)$ nor
the rescattering are needed to describe the data, hence the physical picture
of the two models is rather different. In conclusion, the fit in Ref.~\cite{plb718p1369} is
performed with six parameters (repeated for different values of the
subtraction constant $\mu$, for a total of seven), while we use only 4. Both
approaches can describe data: in the future, better data are needed to
understand which effects are more important. Our work suggests that the nearby
$\bar{D}D$ threshold and $\bar{D}D$ loops are necessary to correctly
understand the $\psi(3770)$. Furthermore, the presence of two poles is a stable
prediction of our approach, see Secs.~\ref{results.2} and \ref{results.3} below.

\bigskip

i) As a last comment, we compare our result with the
approach of Refs.~\cite{prd86p114013,prd87p057502}. Interestingly, in Ref.~\cite{prd87p057502} it is shown that a
suppression of the cross-section at about $3.81$ GeV is obtained by a
destructive interference of the resonance $\psi (3770)$ with the background.
This outcome is useful, since it sets a physical limit for our approach, in which no
background is present: our description of data cannot be expected above this
upper limit (and indeed we perform the fit below it). However, for what
concerns the treatment of $\psi (3770)$ our model is quite different: in
Refs.~\cite{prd86p114013,prd87p057502} rescattering is taken into account. When their rescattering
parameter is set to zero, the real part of the $D\bar{D}$ loop vanishes. In
contrast, in our approach, even without an explicit rescattering term, the $D%
\bar{D}$ loops are taken into account in a way that guarantees the unitarity
expressed by Eq.~\eqref{unit} (its fulfillment is one of the most relevant technical
aspects of our approach) and they are found to be important for a proper
description of the $\psi (3770)$.


\subsection{\label{results.2} Pole positions}

As renowned, a theoretically (and lately experimentally as well, see e.g. the $f_0(500)$ in PDG)
stable approach to describe unstable states is based on the determination of
the position of the corresponding poles. In the present case, one obtains, quite
remarkably, two poles on the III Riemann Sheet. As seen in Sec.~\ref{model.2}, this is the sheet relevant for
a system with two decay channels at energy above both thresholds. In the
isospin limit, the III RS reduces to the usual II RS. In the following, the indeterminacy of the pole masses is estimated to
be $\sim 1.0$ MeV, close to the error obtained for the parameter $m_{\psi}$, and the indeterminacy of the pole width(s) as $\sim 1.0$ MeV,
coincident with the error for $\Gamma_{D\bar{D}}$ (see Table \ref{widths}). In fact, the errors
of these strictly correlated quantities must have a similar magnitude. 
The closest pole to the Riemann axis reads:%
\be
\label{pole1}
\text{First pole: }\ E=3777.0-i12.3\text{ MeV, }%
\ee
hence%
\be
\begin{split}
&m_{\psi}^{\text{pole}}\simeq 3777.0\pm 1.0\text{ MeV and }\\
&\Gamma_{\psi}^{\text{pole} }\simeq 24.6\pm 1.0\text{ MeV,}
\end{split}
\ee
which agrees quite well with the PDG values \cite{pdg}. This pole is closely
related to the {\it seed} charmonium state and to the position and width of the
peak. Furthermore, a second broader pole appears at lower energy:%
\be
\label{pole2}
\text{Second pole: }\ E=3741.1-i18.4\text{ MeV.}%
\ee
This pole is \emph{dynamically generated}, and it is related to the the
deformation of the signal on the left side. It is also referred to as a
companion pole emergent due to the strong dynamics. The situation is very much
reminiscent of the light scalar $\kappa$ state at lower energy, which was interpreted
as a dynamically generated companion pole of $K^*_0$ in Refs.~\cite{npb909p418,apps9p467}. For additional recent
works on this state, see Refs.~\cite{epjc77p431,prd93p074025}, and references therein.
This similarity is interesting also in view of an important
difference between the two systems: the decay of a scalar kaon into two
pseudoscalars is a $s$-wave ($\Gamma\sim k$) while that of a vector charmonium
into two pseudoscalars is a $p$-wave ($\Gamma\sim k^{3}$).

We mention again \ref{C}, where a different form factor is
used. The results turn out to be very similar and also in that case two poles,
very close to the ones discussed here, are found.

No other poles could be numerically found in the complex
plane. 
 As discussed in comment e), the energy scale $\Lambda 
$ does not represent a limit of validly of our approach. Yet, since the
numerical value of $272$ MeV preferred by the fit is rather small (the value
of the emitted momentum $k(m_{\psi }^{2},m_{D^{0}})$ is comparable to it),
it is interesting to investigate if two poles exist also when artificially
increasing its value. Upon fixing $\Lambda =350$ MeV, one gets $\chi
^{2}=35.05,$ hence the description of data is worsened ($\chi ^{2}/d.o.f=3.5$%
) but still qualitatively acceptable; nevertheless, there are still two poles: the seed
one corresponds to $3.770-i0.018$ GeV (somewhat too large) and the broad
dynamically generated one to $3.755-i0.040$ GeV. Upon increasing $\Lambda $ even
further to $400$ MeV, the $\chi ^{2}$ worsens, i.e.~$\chi ^{2}/d.o.f=4.98$
(which is already beyond the border of a satisfactory description of data), but
there are still two poles; the seed pole reads $3.769-i0.016$ GeV and the
companion pole $3.767-i0.062$ GeV. This exercise shows that the presence of
two poles is a stable result also when changing the value of $\Lambda $,
even to values which are safely larger than the value of the modulus of the
outgoing momentum of the $D$ meson (even if the data description gets
worse for increasing $\Lambda $).

\subsection{\label{results.3} Pole Trajectories}

Here we study the effect of varying the coupling constant over the line-shape and over the pole positions. We define $\tilde{g}=\lambda g_{\psi \dd}$ as the variable parameter (upon varying the scaling factor $\lambda$) and let $g\equiv g_{\psi \dd}$ be fixed to the value in Table \ref{fitpar}. In Quantum Chromodynamics (QCD), a change in such coupling is equivalent to vary the number of colors according to the scaling $\lambda\propto N_{c}^{-1/2}$. Results are shown in Fig.~\ref{gvar} for the cross-section, and in Table \ref{t4} for pole positions. The pole trajectories, as a function of $\tilde{g}$, are depicted in Fig.~\ref{traj}. As expected, as $\tilde{g}$ decreases, the cross section approaches a Breit-Wigner-like shape, see case where $\tilde{g}=0.7g$ in Fig.~\ref{gvar}. Figure \ref{traj} shows that, as $\tilde{g}$ gets smaller, the first pole, from the seed, gets closer to the real axis, while the second pole, dynamically generated, moves deeper in the complex plane until it disappears at the $\ddn$ threshold for a certain critical value, in this case $\tilde{g}<26.9$ (corresponding to $N_c=3.9$). On the contrary, when $\tilde{g}$ increases, the first
pole runs away from the real axis, while the second pole approaches it. Eventually, the dynamically
generated pole gets even closer to the real axis than the seed pole, as it is shown for the case where $\tilde{g}=1.3g$. For such a value of $\tilde{g}$ (or larger), the second pole originates a second peak in the line-shape (see Fig.~\ref{gvar}).

\bfig[ptb]
\begin{center}
\resizebox{!}{200pt}{\includegraphics{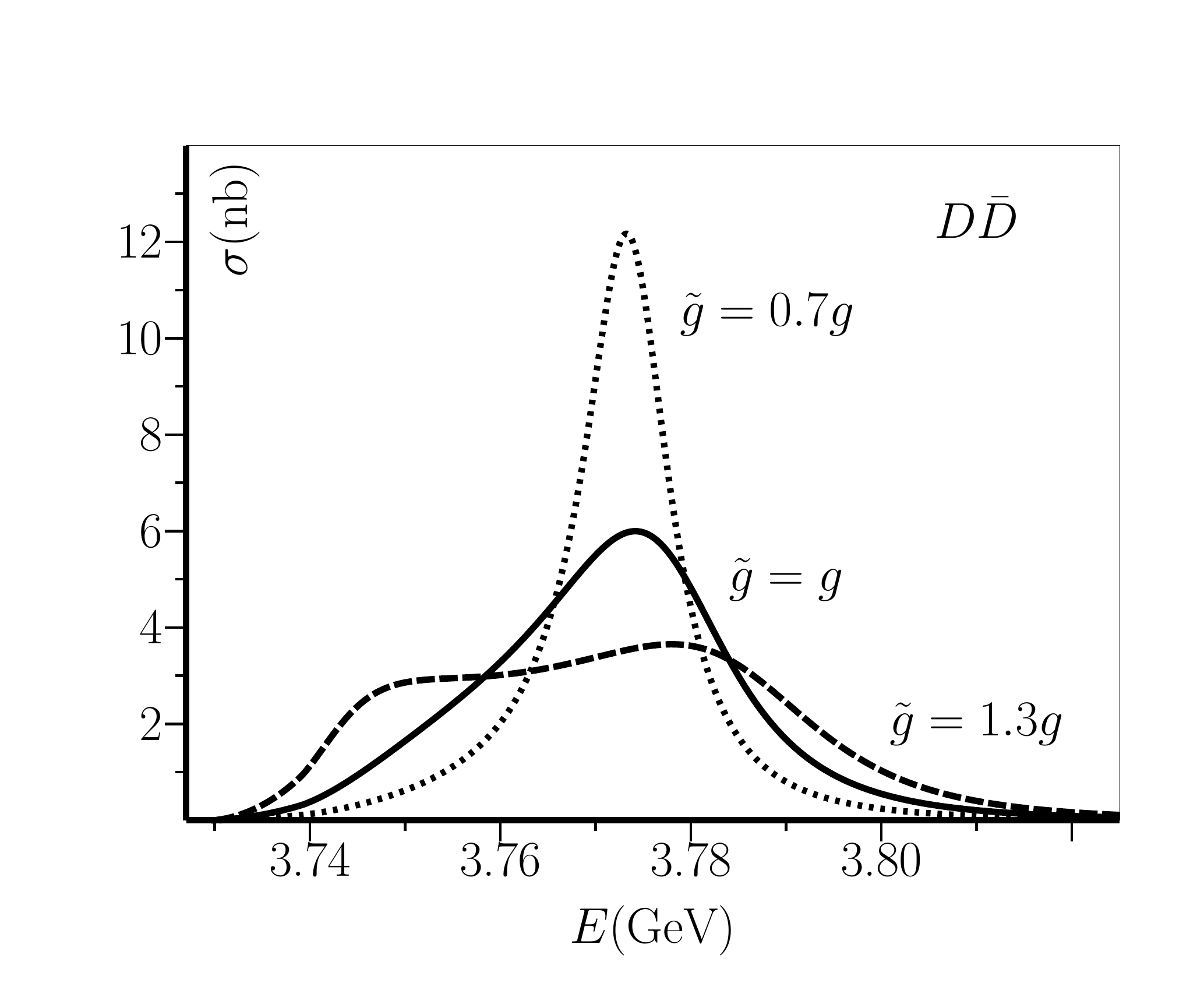}}
\end{center}
\caption{Variation of the cross section in channel $\dd$ with the
coupling $\tilde{g}$. Here, $g\equiv g_{\psi\dd}=30.73$, and the other parameters are found in Table \ref{fitpar}. See corresponding poles in Table \ref{t4}.}%
\label{gvar}%
\efig

\begin{table}[ptb]
\begin{center}%
\btb
[c]{cccc}\hline
$\tilde{g}$ & $0.7{g}$ & $g\equiv g_{\psi\dd}$ 		& $1.3{g}$\\\hline
Pole 1 & $-$ 		& $3741.1-i18.4$ & $3741.0-i9.5$\\
Pole 2 & $3773.7-i5.5$ & $3777.0-i12.3$ & $3785.0-i17.1$\\\hline
\etb
\end{center}
\caption{\label{t4}Variation in the poles position with the coupling $\tilde{g}$, for
parameters in Table \ref{fitpar} (cf.~Fig.~\ref{gvar}).}%
\end{table} 

\bfig[ptb]
\begin{center}
\resizebox{!}{200pt}{\includegraphics{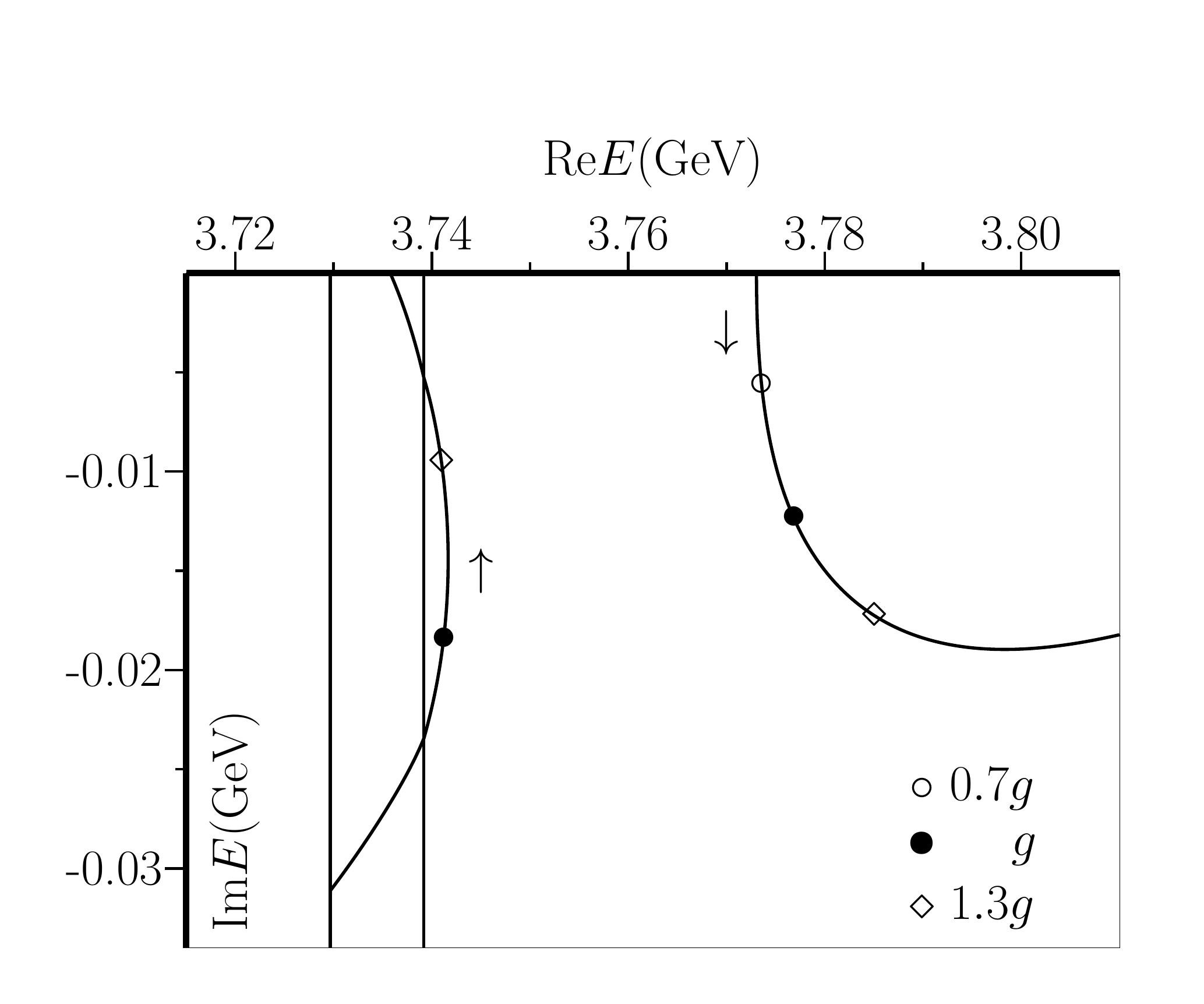}}
\end{center}
\caption{Pole trajectories. Here, $g\equiv g_{\psi\dd}=30.7$, and the other parameters can be found in Table \ref{fitpar}. For the numerical values of the poles at positions $\bullet$, $\diamond$, and $\circ$, see Table \ref{t4}. The vertical lines correspond to the $\ddn$ and $\ddc$ thresholds, from left to right. The arrows show the direction of increasing the coupling (cf.~text).}%
\label{traj}%
\efig

\section{\label{conclusion} Summary, Conclusions, and Perspectives}

Vector mesons can be produced by a single virtual photon, hence they play a crucial role in the experimental study of QCD. The renowned charmonium vector state $J/\psi$ opened a new era in
hadron physics and it is one of the best studied resonances (cf.~\cite{pdg}). Here, we have focused on an orbital excitation of the $J/\psi$, the meson
$\psi(3770)$, whose mass lies just above the $\dd$ threshold. For this
reason, this resonance is extremely interesting. The cross-section
$e^{+}e^{-}\rightarrow\psi\rightarrow \dd$ encloses nonperturbative
phenomena nearby the OZI-allowed open decay channel $\dd$. The production
and decay to all hadrons shows evidence of a deformation on the line-shape of the resonance \cite{prl101p102004}. 

In this work, we presented an unitarized effective
Lagrangian model, with one bare $\bar{c}c$ orbital  seed state $\psi$ dressed by $\dd$
mesonic loops, that has never been employed before to this system. In
particular, our model accounts for one-loop $\ddc+\ddn$
contributions by fulfilling unitarity of the spectral function. Using four
parameters only, we obtain a satisfactory fit to the $\dd$ cross-section
data ($\chi^{2}/$ ${d.o.f.}\sim 1.03$). Moreover, the theoretical cross section is in good agreement with data in channels $\ddc$ and $\ddn$ separately. Partial widths are also in agreement with the PDG values. The
partial decay to leptons is smaller than the average in PDG, but still
reasonable (and in agreement with Ref.~\cite{plb711p292}). The cutoff parameter $\Lambda$,
inversely proportional to the size of the wave-function, turns out to be
rather small, possibly due to the fact that the $\psi(3770)$ is a $D$-wave. Other effects, such as final state rescattering modelled by a four-leg
interaction term, or the contribution of the tail of the $\psi(2S)$ to the
amplitude, are not taken into account, since our fit is already very good
without them (this represents a difference with Ref.~\cite{plb718p1369}, where these terms
were necessary to describe data but where the normalization of the spectral
function of $\psi(3770)$ was not guaranteed). The detailed study of the
role of these additional effects within our framework needs more precise data
and is left for the future.

An important result of our work is the evaluation of two poles at $3777-i12$
and $3741-i19$, the first is a standard seed pole, very close to the one
determined by BES in Ref.~\cite{prl101p102004}, and the second broader pole is dynamically
generated, being an additional companion pole. The presence of a second pole
explains the deviation from a pure Breit-Wigner line-shape. By varying the
coupling constant $g_{\psi D\bar{D}}$, we studied the pole trajectories and
its influence over the line-shape. For small couplings the dynamical pole
disappears and the seed pole approaches the real axis, leading to a
Breit-Wigner-like line-shape, while for larger couplings the dynamical pole
approaches the real axis while the seed pole moves down and to the right, and
a second peak becomes evident (see Figs.~\ref{gvar} and \ref{traj} for
details).} The width of about 24 MeV to channel $\dd$, that we obtain from the first pole, is consistent with the experimental measurements to this channel, and with the branching fraction estimated to be about $85\%$ in Ref.~\cite{plb659p74}, considering that measurements of the width in decays to all hadrons are a few MeV higher (cf.~Ref.~\cite{pdg}).

A natural outlook is the inclusion of the many OZI-suppressed hadronic
decay channels. A related topic is the evaluation of mixing with
the vector glueball, whose mass is predicted to be at about 3.8 GeV by lattice
QCD \cite{prd73p014516}. Finally, the same theoretical approach used in this work can be applied to other resonances whose
nature is not yet clarified, the so-called $X,$ $Y,$ and $Z$ states
\cite{xyz,pr668p1,epja36p189}. It is well possible that some of these
resonances arise as companion poles due to a strong coupling of a standard
charmonium seed state to lighter mesonic resonances. The strong
dressing of the original state with meson-meson clouds generates new poles and
new peaks. An intriguing possibility is that the additional peak arises at a 
higher energy than the seed state, see Refs.~\cite{zpc30p615,prd65p114010,zpc68p647,prd95p034015} for a phenomenological description of this phenomenon in the light sector. Hence, detailed studies of the dynamics underlying each line-shape, that consider poles and interference with the main hadronic degrees of freedom, may shed light on some newly discovered resonances in the charmonium region.

An interesting outlook should be the
description of the whole sector $1^{--}\ $above the $D\bar{D}$ threshold. To
this end, one should develop a unique treatment of the four standard
charm-anticharm resonances above the $D\bar{D}$ threshold $\psi (3770)$, $%
\psi (4040)$, $\psi (4160)$, and $\psi (4415)$ (and also the resonance $\psi (2S)
$ below). In this way, following and extending the discussion in Refs.~\cite{prd94p014016,plb718p1369}, all quantum mixing terms should be properly taken into account. The
aim of such an ambitious project would be the simultaneous description of
the whole sector as well as the description of (some of the) newly observed $%
Y$ resonances in this energy region, which might emerge as dynamically
generated states.

\section*{Acknowledgements}

We thank G.~Rupp and E.~van Beveren for useful discussions. This work was supported by the
\textit{Polish National Science Center} through the project OPUS no.~2015/17/B/ST2/01625.

\appendix


\section{\label{A}Full Lagrangian}

The complete form of the Lagrangian used in this work is given by:%

\begin{align}
\mathcal{L}_{0} &  =-\frac{1}{4}V_{\mu\nu}V^{\mu\nu}+\frac{1}{2}m_{\psi
}^{2}\psi_{\mu}\psi^{\mu}
  +\frac{1}{2}\Big(\ \partial_{\mu}D^{+}\partial^{\mu}D^{-}-m_{D^{+}}%
^{2}\ddc\Big)\\
&  +\frac{1}{2}\Big(\ \partial_{\mu}D^{0}\partial^{\mu}\bar{D}^{0}-m_{D^0}%
^{2}D^0\bar{D}^{0}\Big)
  +\sum_{l=e,\mu,\tau}\bar{\Psi}_{l}(i\gamma^{\mu}\partial_{\mu
}-m_{l})\Psi_{l}\ ,
\end{align}
\be
\label{lagt}
\mathcal{L}_{full}=\mathcal{L}_{0} +\mathcal{L}_{\psi \dd}+\mathcal{L}_{\psi l^{+}l^{-}}\text{ ,}%
\ee
where in the first line the definition $V_{\mu\nu}=\partial_{\mu}\psi_{\nu
}-\partial_{\nu}\psi_{\mu}$ has been implemented, and the interaction terms in Eq.~\eqref{lagt} are given by Eqs.~\eqref{lagi} and \eqref{lagii}.


\section{\label{B}Details of the Loop Function}

Here, we derive the propagator of the vector resonance $\psi$ dressed by
$\dd$ loops. The considerations are however general, and they can be applied
to loops of each vector state. Using the shortening notations%

\begin{align}
&\int_{q}\equiv\int\frac{d^{4}q}{(2\pi)^{4}}\text{ ,}\\
&\mathcal{D}\equiv[(q+\frac{p}{2})^{2}-m^{2}+i\varepsilon][(q-\frac{p}{2})^{2}-m^{2}%
+i\varepsilon]\ ,
\end{align}
we decompose the tensor loop contribution $\Sigma_{\mu\nu}(p,m)$ in its transverse and longitudinal parts (see e.g.~Ref.~\cite{ppnp39p201}):

\be
\Sigma_{\mu\nu}(p,m)=i\int_{q}\frac{\ 4q_{\mu}q_{\nu}f_{\Lambda}^{2}%
(\xi)}{\mathcal{D}}
=\left(  -g_{\mu\nu}+\frac{p_{\mu}p_{\nu}}{p^{2}}\right)
\Sigma+p_{\mu}p_{\nu}\Sigma_{L}\text{ .}\nonumber
\ee
Then, multiplying by $g^{\mu\nu}$and $p^{\mu}p^{\nu}$, respectively, we get:%

\begin{align}
&g^{\mu\nu}\Sigma_{\mu\nu}(p,m) =i\int_{q}\frac{\ 4q^2f_{\Lambda}^{2}(\xi)}{\mathcal{D}}=-3\Sigma+p^{2}\Sigma_{L}\text{ ,}\\
&p^{\mu}p^{\nu}\Sigma_{\mu\nu}(p,m)=i\int_{q}\frac{\ 4(p\cdot
q)^{2}f_{\Lambda}^{2}(\xi)}{\mathcal{D}}=p^{4}\Sigma_{L}\text{ ,}%
\end{align}
out of which we find, in the rest-frame of the decaying particle,
{$p=(E,\vec{0})$}, where $E$ is the running mass of the $\psi(3770)$:%
\begin{align}
\label{b5}
\Sigma(s,m) &  =\frac{1}{3}\left(  -g^{\mu\nu}+\frac{p^{\mu}p^{\nu}}{p^{2}%
}\right)  \Sigma_{\mu\nu}(p,m)\\
&  =-\frac{4}{3}i\int_{q}\frac{\ \vec{q}^2f_{\Lambda}^{2}(\xi)}{\mathcal{D}%
}\\%
&  =-\frac{1}{2\pi^2}\int_0^\infty d|{\bf q}|\frac{\ \frac{4}{3}{\bf q}^4f_{\Lambda}^{2}(\xi=4({\bf q}^2+m^2))}{\sqrt{{\bf q}^2+m^2}(\xi-s)}\text{ .}%
\end{align}
Namely, its imaginary part implies that $\left\vert \mathbf{q}\right\vert $ is
replaced by $\mathbf{k}_{f}$ and $\xi$ by $s$. Explicitly, it reads:
\be
\label{b7}
\operatorname{Im}\Sigma(s,m)=\sqrt{s}\frac{4}{3}\frac{\left\vert \vec{k}%
_{f}\right\vert ^{3}}{8\pi s}f_{\Lambda}^{2}(s)=\sqrt{s}\Gamma(s,m)\ ,
\ee
as it should. Similarly, for the sum of the two contributions, one has:%
\begin{align}
&\Pi(s)=\frac{1}{3}\left(  -g^{\mu\nu}+\frac{p^{\mu}p^{\nu}}{p^{2}%
}\right)  \Pi_{\mu\nu}(p,m)\text{ ,}\\
&p^{\mu}p^{\nu}\Pi_{\mu\nu}(p,m) =p^{4}\Pi_{L}\text{ .}%
\end{align}
Indeed, an explicit resummation of the full propagator shows that%
\be
\begin{split}
&\Delta_{\mu\nu}(p)=G_{\mu\nu}(p)+G_{\mu\mu^{\prime}}%
(p)\Pi_{\mu^{\prime}\nu^{\prime}}(p)G_{\nu^{\prime}\nu}(p)+\cdots\\
&  =G_{\mu\nu}(p)+\left(  \frac{-g_{\mu\mu^{\prime}}+\frac{p_{\mu}%
p_{\mu^{\prime}}}{p^{2}}}{p^{2}-m_{\psi}^{2}}-\frac{p_{\mu}p_{\nu}}%
{p^{2}m_{\psi}^{2}}\right)\times\\
&  \left[  \left(  -g_{\mu^{\prime}\nu^{\prime}%
}+\frac{p_{\mu^{\prime}}p_{\nu^{\prime}}}{p^{2}}\right)  \Pi%
+p_{\mu^{\prime}}p_{\nu^{\prime}}\Pi_{L\text{ }}\right]
  \left(
\frac{-g_{\nu^{\prime}\nu}+\frac{p_{\nu^{\prime}}p_{\nu}}{p^{2}}}%
{p^{2}-m_{\psi}^{2}}-\frac{p_{\nu^{\prime}}p_{\nu}}{p^{2}m_{\psi}^{2}}\right)
+\cdots\\
&  =\frac{-g_{\mu\nu}+\frac{p_{\mu}p_{\nu}}{p^{2}}}{p^{2}-m_{\psi}^{2}}%
-\frac{p_{\mu}p_{\nu}}{p^{2}m_{\psi}^{2}}+\frac{-g_{\mu\nu}+\frac{p_{\mu
}p_{\nu}}{p^{2}}}{\left(  p^{2}-m_{\psi}^{2}\right)  ^{3}}\Pi
-\frac{p_{\mu}p_{\nu}}{p^{2}M_{0}^{2}}\frac{\Pi_{L}}{m_{\psi
}^{2}}+\cdots\\
&  =\frac{-g_{\mu\nu}+\frac{p_{\mu}p_{\nu}}{p^{2}}}{p^{2}-m_{\psi}^{2}%
+\Pi}-\frac{p_{\mu}p_{\nu}}{m_{\psi}^{2}-\Pi_{L}^{2}}\text{ ,}%
\end{split}
\ee
where we have used%
\be
G_{\mu\nu}(p)=\frac{-g_{\mu\nu}+\frac{p_{\mu}p_{\nu}}{M_{0}^{2}}}%
{p^{2}-m_{\psi}^{2}}=\frac{-g_{\mu\nu}+\frac{p_{\mu}p_{\nu}}{p^{2}}}%
{p^{2}-m_{\psi}^{2}}-\frac{p_{\mu}p_{\nu}}{p^{2}m_{\psi}^{2}}\text{ .}%
\ee
The final result reads%
\be
\Delta_{\mu\nu}(p)=\frac{-g_{\mu\nu}+\frac{p_{\mu}p_{\nu}}{p^{2}}}%
{p^{2}-m_{\psi}^{2}+\Pi}-\frac{p_{\mu}p_{\nu}}{m_{\psi}^{2}-\Pi
_{L}^{2}}\text{ ,}%
\ee
where it is evident that the first part contains the properties of the unstable state.


\section{\label{C} Using a different Cutoff Function}

The use of an exponential vertex function is a typical choice in hadron
physics. (It is also used as a form factor in various phenomenological
approaches \cite{prd32p189,prd53p295}. Yet, for completeness, we test here a different
cutoff function, chosen as:
\begin{equation}
f_{\Lambda}(\xi)=\frac{1}{1+\left[  \frac{\xi^{2}-2(m_{D^{0}}+m_{D^{+}})^{2}%
}{4\Lambda^{2}}\right]  ^{2}}\ ,\label{cutoffnew}%
\end{equation}
where $\xi=4(\mathbf{q}^{2}+m^{2}),$ $m$ being the mass of the particle
circulating in the loop. In the isospin limit the vertex function reduces to
$\left(  1+\mathbf{q}^{4}/\Lambda^{4}\right)  ^{-1}.$ Since in the loop
integral the squared function $f_{\Lambda}^{2}(\xi)$ enters, convergence is
guaranteed (see also Eqs.~\eqref{b5}-\eqref{b7}). Notice that a simple dipole function of the type $\left(  1+\mathbf{q}^{2}/\Lambda^{2}\right)  ^{-1}$ would not fall fast enough.

Clearly, also with Eq.~\eqref{cutoffnew} the loop function expressed in Eq.~\eqref{eq21} is, besides a cut on the real axis, regular everywhere in the first
Riemann sheet.
In the second Riemann sheet of $\Sigma(z^{2},m)$, it enters the vertex function $f_{\Lambda}(z^{2})$,
continued to the complex plane, and
additional poles emerge.

Let us now present the results. By performing the very same fit as in Sec.~\ref{results},
we find a minimum for $\chi^{2}/d.o.f.=9.35/10$, which is even smaller than the one reported in Table \ref{fitpar}. The corresponding cutoff $\Lambda=266.4$
MeV, mass $m_{\psi}=3773.39$, and coupling to leptons $g_{\psi e^{+}e^{-}%
}=1.054\times 10^{-3}$ are very similar to those in Table \ref{fitpar}. Only the
coupling constant turns out to be slightly smaller,  $g_{\psi D\bar{D}}=23.9$. However, the tree-level decay widths $\Gamma_{D^{0}\bar
{D}^{0}}=11.4$ MeV and $\Gamma_{D^{+}D^{-}}=7.9$ MeV  are basically unchanged.
Also the spectral function and the cross-sections have a very similar
form, hardly distinguishable to the ones presented in Sec.~\ref{results}.

Finally, we turn to the poles. Also in this case two poles are found, the
first at $3777.2-i11.2$ MeV, very similar to Eq.~\eqref{pole1}, and the
second at $3745.9$ $-i14.7$ MeV, which corresponds well to Eq.~\eqref{pole2}. The
presence of two poles is stable against parameter variations, in a way very
much similar to Fig.~\ref{traj}.

This exercise shows that the precise form of the vertex function
does not affect the results as long as it falls off sufficiently fast. On the
other hand, the numerical value of the cutoff energy scale is crucial. 

\section{\label{D} Three-time subtracted dispersion relation}

As seen in the main text through the Gaussian vertex function, and in the
previous Appendix through a dipole-like function, an energy scale of about $%
\Lambda \approx 250$-$300$ MeV emerges from fits to data. Yet, it is
interesting to test a completely different strategy: a three-time subtracted
dispersion relation, in which no energy scale $\Lambda $ appears (namely,
one needs to subtract at least three times to guarantee convergence, see
below). The explicit form of the scalar part of the propagator of the meson $%
\psi (3770)$ reads:%
\begin{eqnarray}
\Delta (s) &=&A_{1}\left[ s-m_{\psi }^{2}+A_{2}\left( s-m_{\psi }^{2}\right)
^{2}+g_{\psi D\bar{D}}^{2}(R_{\psi D^{0}\bar{D}^{0}}^{\text{3sub}%
}(s)+R_{\psi D^{+}D^{-}}^{\text{3sub}}(s)-A_{0})\right.   \nonumber \\
&&\left. +ig_{\psi D\bar{D}}^{2}\left( \Gamma _{\psi \rightarrow D^{0}\bar{D}%
^{0}}^{\text{local}}(s)+\Gamma _{\psi \rightarrow D^{+}D^{-}}^{\text{local}%
}(s)\right) \right] ^{-1}\text{ ,}
\end{eqnarray}%
where $A_{0},$ $A_{1},$ and $A_{2}$ are a convenient rewriting of the three
subtraction constants. Moreover, 
\begin{equation}
\Gamma _{\psi \rightarrow D^{0}\bar{D}^{0}}^{\text{local}}(s)=g_{\psi D\bar{D%
}}^{2}\frac{k^{3}(s,m_{D^{+}})}{6\pi s}\text{ ; }\Gamma _{\psi \rightarrow
D^{+}D^{-}}^{\text{local}}(s)=g_{\psi D\bar{D}}^{2}\frac{k^{3}(s,m_{D^{+}})}{%
6\pi s}
\end{equation}%
are the tree-level decay widths obtained from the Lagrangian of Eq.~\eqref{lagi}
(without any form factor, hence valid in the local limit), and the real
parts read 
\begin{eqnarray}
\text{ }R_{\psi D^{0}\bar{D}^{0}}^{\text{3sub}}(s) &=&s^{3}\frac{PP}{\pi }%
\int_{4m_{D^{+}}^{2}}^{\infty }\frac{\sqrt{s}\Gamma _{\psi \rightarrow D^{0}%
\bar{D}^{0-}}^{\text{local}}(s)}{(s^{\prime }-s)s^{\prime 3}}\mathrm{ds}%
^{\prime }\text{ , } \\
R_{\psi D^{+}D^{-}}^{\text{3sub}}(s) &=&s^{3}\frac{PP}{\pi }%
\int_{4m_{D^{+}}^{2}}^{\infty }\frac{\sqrt{s}\Gamma _{\psi \rightarrow
D^{+}D^{-}}^{\text{local}}(s)}{(s^{\prime }-s)s^{\prime 3}}\mathrm{ds}%
^{\prime }\text{ }
\end{eqnarray}%
(for $s$ below threshold the integrals are real and the principal part $PP$
is omitted). These integrals are convergent in due to the factor $s^{\prime
3}$ in the denominator of the integrand.

The constant $A_{0}$ is easily obtained as $A_{0}=R_{\psi D^{0}\bar{D}^{0}}^{%
\text{3sub}}(m_{\psi }^{2})+R_{\psi D^{+}D^{-}}^{\text{3sub}}(m_{\psi }^{2})$
in such a way that $m_{\psi }^{2}$ is the zero of $\R\ \Delta (s)^{-1}$%
. The constant $A_{1}$ can be used to guarantee the normalization of the
spectral function.

We then performed a fit to data involving the following four parameters: $%
m_{\psi }$, $g_{\psi D\bar{D}}$, $A_{2}$, and $g_{\psi e^{+}e^{-}}$. Roughly
speaking, the constant $A_{2}$ replaces the cutoff $\Lambda $. The results
for this set-up are not satisfactory. One obtains various local minima for
the $\chi ^{2}$ of the order of $220$ (hence, $\chi ^{2}/d.o.f\gtrsim 22$,
which in practice means that no reasonable description of data is obtained).

Notice that the three-time subtracted dispersion relation is not equivalent,
within our treatment, to a hard cutoff. Namely, such a hard cutoff would
imply that the spectral function $d_{\psi(3770)}(m)$ is proportional to
$\theta(\mathbf{q}^{2}-\Lambda^{2}).$ In this way the normalization condition
of Eq.~\eqref{unit} would still be fulfilled (at the price of having a rather
unphysical behavior of the spectral function, which goes abruptly to zero
above a certain energy threshold). However, such a procedure is clearly
different from the here presented dispersion relation, in which the imaginary
part of the propagator is nonzero for any value of $\sqrt{s}$ above the
threshold. An equivalence of the treatments can be achieved only when
considering the cutoff $\Lambda$ as a very large number: but, as we explained
in the text, we regard the cutoff function as a physical quantity describing
the finite dimension of mesons and their interactions (in fact, a similar
cutoff function emerges in the context of the microscopic $^{3}P_{0}$ model)
and we do not consider the limit $\Lambda\rightarrow\infty.$

Moreover, even if a hard cutoff is not physical for the treatment of the
$\psi(3770)$, it was used in various studies and is also capable to generate
additional poles, as long as its value is finite and of the same order of (or
not much larger than) the energy scale of the system (see Ref.~\cite{prc76p065204} and Refs.~therein). In fact, when $\Lambda$ is finite, the real part is not a simple
constant and additional nontrivial features of the propagator appear.

In conclusion, the negative result of the three-time subtracted dispersion relation is
nevertheless instructive, since it shows that a form factor which takes into
account the finite dimensions of mesons and of their interactions seems to
represent a rather physical feature needed for a correct description of $D%
\bar{D}$ loops.

\end{document}